# Effect of a weak magnetic field on ductile–brittle transition in micro-cutting of single-crystal calcium fluoride

Yunfa Guo[1], Yan Jin Lee[1], Yu Zhang[1], Anastassia Sorkin[1], Sergei Manzhos[2,*] and Hao Wang[1,*]

[1] *Department of Mechanical Engineering, National University of Singapore, 9 Engineering Drive 1, Singapore 117575, Singapore*

[2] *School of Materials and Chemical Technology, Tokyo Institute of Technology, Ookayama 2-12-1, Meguro-ku, Tokyo 152-8552, Japan*

* Corresponding authors. Emails: mpewhao@nus.edu.sg (H. Wang); manzhos.s.aa@m.titech.ac.jp (S. Manzhos)

**Abstract**

Magneto-plasticity occurs when a weak magnetic field alters material plasticity and offers a viable solution to enhance ductile-mode cutting of brittle materials. This study demonstrates the susceptibility of non-magnetic single-crystal calcium fluoride ($CaF_2$) to the magneto-plastic effect. The influence of magneto-plasticity on $CaF_2$ was confirmed in micro-deformation tests under a weak magnetic field of 20 mT. The surface pile-up effect was weakened by 10–15 nm along with an enlarged plastic zone and suppressed crack propagation under the influence of the magnetic field. Micro-cutting tests along different crystal orientations on the (111) plane of $CaF_2$ revealed an increase in the ductile–brittle transition of the machined surface with the aid of magneto-plasticity where the largest increase in ductile–brittle transition occurred along the $[11\bar{2}]$ orientation from 512 nm to a range of 664–806 nm. Meanwhile, the subsurface damage layer was concurrently thinner under magnetic influence. An anisotropic influence of the magnetic field relative to the single-crystal orientation and the cutting direction was also observed. An analytical model was derived to determine an orientation factor *M* that successfully describes the anisotropy while considering the single-crystal dislocation behaviour, material fracture toughness, and the orientation of the magnetic field. Previously suggested theoretical mechanism of magneto-plasticity via formation of non-singlet electronic states in defected configurations was confirmed with density functional theory calculations. The successful findings on the influence of a weak magnetic field on plasticity present an opportunity for the adoption of magnetic-assisted micro-cutting of non-magnetic materials.

## 1 Introduction

The application of the magnetic field in machining processes has been proven beneficial to machining processes as reported in various metal cutting studies. El Mansori and Mkaddem [1] applied an external magnetic field during orthogonal cutting of AISI 1045 steel and found that cutting-induced surface plastic deformation was severely affected by the magnetic field. A harder machined surface and thinner secondary deformation zone at the tool-chip interface were reported where a relationship was established between the rate of surface hardening rate and the secondary zone thickness and the amplitude of magnetic field. Mkaddem and El Mansori [2] later reported that the magnetic field extended the slip band area within the primary shear zone in the cutting tests of AISI 1045 steel, which suggested an improvement in the plasticity of the steel. Dehghani et al. [3] conducted the turning tests on steel alloy 30CrNiMo8 with a magnetic field and ferreted out a remarkable reduction in flank wear of the cutting tool, lower cutting forces, and more continuous chip formations. The observations were attributed to the enhanced material flow and plastic deformation rate induced by the enhanced dislocation motion under the magnetic field.

These early reports on the favourable effects of applying a magnetic field in the cutting process were mainly focused on cutting the ferromagnetic steel alloys with a strong magnetic field. The challenge in studying this combination of experimental conditions is the difficulty in evaluating the effects of the magnetic field in the improvement of cutting performance. Other magnetic effects remain prevalent when studying ferromagnetic materials such as magnetization and magnetostriction. The employment of a strong magnetic field also poses a potential threat to the general safety of the working environment for both machine tools and the human operator's health.

The alteration of material plasticity under a magnetic field has been reported across an even wider range of materials beyond metals, such as semiconductors, and dielectrics with paramagnetic obstacles [4,5]. This is defined as the magneto-plastic effect, which differs from magnetization and magnetostriction effects where the effect is also imposed on non-magnetic materials. Investigations on the magneto-plastic effect date back to the 1970s when Kaganov et al. [6] observed a relationship between electron drag and dislocation motion under a magnetic field. Alshits et al. [7] coined the phenomenon as the magneto-plastic effect by revealing the impact of a constant magnetic field on the dislocation motion of NaCl. Urusovskaya et al. [8] reported a reduction of yield stress during compression tests of LiF, NaCl, and PbS crystals under a weak magnetic field. Molotskii [4] and Alshits et al. [5] revealed that diamagnetic crystals with paramagnetic centres in structural defects were essential to realising the improvement in plasticity under a weak magnetic field. The weak magnetic field is generally specified to the intensities less than 1 T from prior reports [9]. It is important to note that the magneto-plastic effect manifests with weak magnetic field intensities, defined as field intensities less than 1 T [9], as compared to previous applications of stronger field intensities to alter material properties.

This paper will study the influence of a weak magnetic field in the domain of micro-cutting of a non-magnetic insulator material as compared to previous studies of metal cutting on the macro-scale. A non-magnetic brittle material, single-crystal calcium fluoride ($CaF_2$), was studied to focus on magnetic effects other than ferromagnetism or eddy currents which previously were suggested as causing the magneto-plastic effect in non-ferromagnetic materials. $CaF_2$ is widely employed in the optics and semiconductor industries due to its high permeability and outstanding laser-damage thresholds [10]. As $CaF_2$ is a soft-brittle material with low fracture toughness, low hardness, and high thermal expansion coefficient [11], it is challenging to cut crack-free surfaces during ultra-precision machining (UPM). Hence, the identified modifications to the material plasticity under the influence of a magnetic field have a potentially positive impact on augmenting the cutting process. Several other methods to enhance ductile-mode cutting of $CaF_2$, such as thermal-assisted machining [12], ultrasonic vibration-assisted machining [13], and solidified coating [14] were previously used. Instead, this paper will provide supporting evidence of the possibility of employing the magneto-plastic effect as an alternative approach to support plasticity during cutting.

The plasticity of the $CaF_2$ was evaluated by assessing the anisotropic ductile–brittle transition during micro-scale plunge-cutting and further material characterization through micro-indentation tests, machine surface and subsurface assessment. In addition, this work will also uncover the orientation dependence of the magnetic field relative to the anisotropic crystallography [15,16], based on the notion that the magneto-plasticity only work when dislocation slip direction is consistent with the field direction [8,17]. A theoretical model was also used to connect the magneto-plastic effect with the application of machining; in particular by establishing the fulfilment of prerequisites for the manifestation of the magneto-plastic effect using density functional theory calculations and deriving an analytical model to describe the changes in mechanical properties of the material.

**2 Magneto-plasticity in CaF$_2$**

*2.1 Dislocation depinning in CaF$_2$*

Golovin [9,18] attributed the manifestation of the magneto-plastic effect to the theory of spin-dependent dislocation depinning where a magnetic field promotes the transformation of radical pairs between dislocations and stoppers from a singlet state to a triplet state that possesses lower binding energy. To trigger the spin conversion of radical pairs under a magnetic field, the stopper must involve paramagnetic obstacles, which include point defects with unpaired electrons [4].

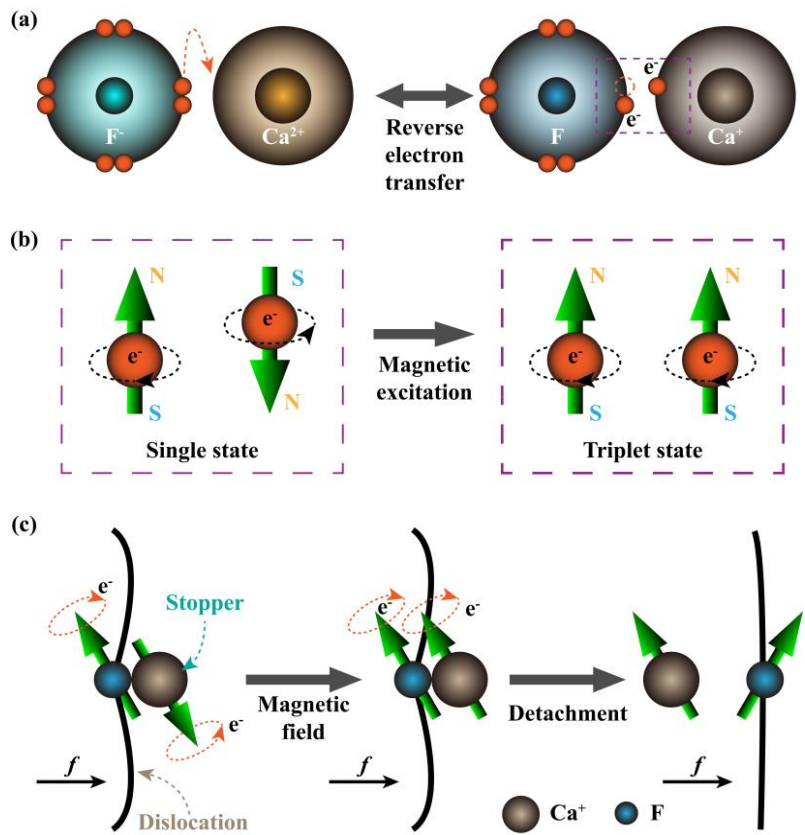

**Fig. 1.** (a) Reverse electron transfer from a F$^-$ ion to a Ca$^{2+}$ ion; (b) Spin conversion of radical pairs under the magnetic excitation; (c) Detached dislocations from stoppers in a magnetic field.

Four potential intrinsic lattice defects are known in CaF$_2$ single crystals as shown in Fig. S1, which include Ca and F interstitials, Ca and F vacancies, Frenkel and anti-Frenkel defects [19]. There are existing defects in the singlet ground state but possess triplet excited states (e.g., Ca vacancies and interstitials), and others that are non-singlet and already in the ground state (e.g., F vacancies and interstitials). Other defects that can form during plastic deformation involve the displacement of atoms such as vacancies, stacking faults, and dislocation emission. These defects may also be non-singlet in the ground state or possess sufficiently low-lying non-singlet states that can interact with an external magnetic field. In particular, it has been noted that between dislocations and stoppers, a reverse electron transfer can occur, which creates unpaired electrons for the dislocations and stoppers [20] that can interact with a magnetic field. Fig. 1(a) illustrates such an occurrence where an electron is transferred from an $F^-$ ion to a $Ca^{2+}$ ion, and a new pair (F $Ca^+$) with unpaired electrons is generated. Therefore, the $CaF_2$ single crystals with paramagnetic point defects qualify for potential dislocation depinning under the influence of a magnetic field.

In the absence of external excitation, the radical pairs are in the singlet state with opposite spin directions. The presence of magnetic excitation then arguably energises the radical pairs to triplet states where the spin direction of radical pairs is uniform and in alignment with the direction of the external magnetic field (Fig. 1(b)). Fig. 1(c) illustrates the detachment of a dislocation from a stopper under the application of magnetic field, where the radical pair (F $Ca^+$) is assumed as a "dislocation + stopper" system. Under conventional conditions, the dislocation (F) is pinned by the stopper ($Ca^+$) since the radical pair ($F^-$ $Ca^+$) is in the single state with strong binding energy and the applied stress $f$ is inferior to the critical pinning force $f'$ between the dislocation (F) and the stopper ($Ca^+$). The radical pair (F $Ca^+$) would be excited to the triplet state under a magnetic field, which significantly reduces the binding energy and the critical pinning force of the "dislocation + stopper" system [21].

*2.2 First-principle calculations*

Density functional theory (DFT) calculations of defect formation energies, ground-state spin states, and singlet-triplet gaps were performed for different types of defects to establish the fundamental pre-requisites for the work material to be affected by the magneto-plastic effect. This study encompasses the introduction of a stacking fault and the combined influence of dislocations into the simulation cells. These conditions were created by cutting out a row of $CaF_2$ units along an axis of the cell. The full details of the computational approach, the defects considered, and the results are given in the Supporting Information.

Most defects have rather high formation energies on the order of 10 eV, which is in agreement with previous reports [22,23] (see Table S1). These defects are unlikely to form under normal conditions but could be induced under cutting conditions. The formation energies and dislocations and stacking faults as well as (non-singlet) F interstitials were computed to be low enough for these types of defects to be induced at normal conditions. Some defects (including doublet F interstitials and vacancies, some double vacancies which are triplet or quintet, and the doublet Ca vacancy, as compiled in Table S1) are non-singlet and can directly interact with the external magnetic field. For most defects that are singlet in the ground state, the singlet-triplet gap is of several eV, which implies that triplet states cannot be induced by the external field. However, the singlet-triplet gaps for the stacking fault and two dislocations with a stacking fault between them are low, on the order of room-temperature $kT$, and these systems are likely to be excited into triplet states under normal conditions. These defects also possess low formation energies of less than 0.1 eV/atom. Therefore, these types of defects can enable the dislocation-depinning mechanism of weakening the pinning force as described in Section 2.1, which supports the possibility of observing the magneto-plastic effect in $CaF_2$ as suggested in the literature.

*2.3 Magneto-plastic effect on mechanical properties*

Having established the likelihood of observing the magneto-plastic effect in $CaF_2$, this section presents the derivation of a model to describe the influence of the magnetic field on the mechanical properties of the material.

Based on the dislocation-depinning theory, applied stress would be capable of easily detaching the dislocation from the stopper as a result of the weakened critical pinning force. Dislocation motion is then assumed to occur when the external force $F$ acting on a unit length of dislocation exceeds a critical value $F'$ as determined with Eq. (1) [21]:

$$F' = b\sigma_i \approx \sqrt{\frac{(\gamma f')^3}{bG} C_V} \tag{1}$$

where $b$ is the magnitude of Burgers vector, $\sigma_i$ is the internal stress acting on dislocation, $G$ is the shear modulus, $C_V$ is the volume concentration of point defects, and $f'$ is the critical pinning force that characterizes the dislocation detachment from stoppers. From Eq. (1), a magnetic field promotes dislocation mobility by reducing the critical pinning force through the radical pair transformation process.

The velocity of magnetic field-enhanced dislocation movement is proportional to the square of the external magnetic field intensity in a weak magnetic field [17,24]:

$$v_B = v_0 \left(1 + k\frac{B^2}{B_0^2}\right) \tag{2}$$

where $v_B$ and $v_0$ are the velocities of dislocations in the presence and absence of a magnetic field, $B$ is the intensity of the applied magnetic field, $B_0$ is a characteristic magnetic field constant needed to induce depinning of the dislocations from paramagnetic centres, and $k$ is a proportionality factor.

An increase in velocity of dislocations will increase the dislocation path length $L$, which is correlated with the total dislocation density $\rho_t$ in a crystal that is defined as [21]:

$$L \sim \frac{1}{\sqrt{\rho_t}} = \frac{1}{\sqrt{\rho_{fr} + \rho_{old}}} \tag{3}$$

where $\rho_{fr}$ is the fresh dislocation density produced by external excitation and $\rho_{old}$ is the pre-excitation dislocation density. A high density of dislocations would increase interaction and entanglement between dislocations, which would affect the mechanical strength of the material. Therefore, the change in the dislocation path length $L$ can also be assumed to be proportional to $B^2$ [4], and the correlation between the total dislocation density and external magnetic field in a weak magnetic field is derived as follows:

$$\rho_{tB} = \rho_{t0} / \left(1 + k_\rho \frac{B^2}{B_0^2}\right)^2 = \rho_{t0} \left(\frac{B_0^2}{B_0^2 + k_\rho B^2}\right)^2 \tag{4}$$

where the total dislocation density in the presence and absence of a magnetic field are assigned as $\rho_{tB}$ and $\rho_{t0}$, respectively, $k_\rho$ is a proportionality factor with respect to the total dislocation density.

The mechanical hardness $H$ is defined by the dislocation density as follows [25]:

$$H = C\zeta Gb\sqrt{\rho_s + \rho_g} \approx C\zeta Gb\sqrt{\rho_t} \tag{5}$$

where $C$ is a constraint factor, $\zeta$ is the material constant, $\rho_s$ is the statistically stored dislocations (SSDs), $\rho_g$ is the geometrically necessary dislocations (GNDs), and $\rho_t$ is the total dislocation density (i.e., the sum of SSDs and GNDs). The relative change of hardness $H$ in a weak magnetic field can then be defined as a function of the magnetic field intensities by substituting Eq. (4) into Eq. (5) as follows:

$$\frac{H_B}{H_0} = \frac{B_0^2}{B_0^2 + k_\rho B^2} \tag{6}$$

where $H_B$ and $H_0$ are material hardness with and without a magnetic field.

As this work evaluates the ductile–brittle transition of a brittle material, the fracture toughness material parameter $K_c$ is an important factor to consider. Based on Griffith's understanding of fracture mechanics [26], the increase in fracture toughness elevates the resistance to crack formation. The values of $K_c$ can be estimated using the material hardness and the length of cracks formed during micro-indentation tests as follows [27]:

$$K_c = 0.016 \left(\frac{E}{H}\right)^{0.5} \frac{P}{c^{1.5}} \tag{7}$$

where $E$ is Young's modulus, $P$ is the applied load, and $c$ represents the average distance from the indentation centre to the tip of the propagated cracks. Consequently, the fracture toughness under the influence of a magnetic field is determined by substituting Eq. (6) into Eq. (7) to derive the change in fracture toughness as a function of the magnetic field intensity as follows:

$$\frac{K_{cB}}{K_{c0}} = \left(\frac{H_0}{H_B}\right)^{0.5} \left(\frac{c_0}{c_B}\right)^{1.5} = \left(1 + k_\rho \frac{B^2}{B_0^2}\right)^{0.5} \left(\frac{c_0}{c_B}\right)^{1.5} \tag{8}$$

where $K_{cB}$ and $K_{c0}$ are the fracture toughness with and without a magnetic field, and $c_B$ and $c_0$ is the crack length with and without a magnetic field.

*2.4 Anisotropic magneto-plastic effects during micro-cutting.*

The orientation-dependence of the magneto-plastic effect should be considered based on previous observations that the magnetic field cannot militate in favour of plastic deformation when the dislocation glide is vertical to the magnetic field direction [8,17]. The magnetic field reduces the required force for dislocation motion as a function of the magnetic field intensity as described by Eq. (9). Orientation dependence of the magnetic field lines relative to the dislocation motion also exists [17] such that a field orientation parallel to the dislocation motion promotes the magneto-plastic effect. Hence, the effective magnetic field intensity is determined by Eq. (10).

$$F_B \propto B_e^2 \tag{9}$$

$$B_e^2 = B^2 \cos\theta \cos\lambda \tag{10}$$

where $\theta$ is the angle between magnetic field direction and dislocation slip plane normal and $\lambda$ is the angle between magnetic field direction and dislocation slip direction.

As the magnetic effect on dislocation motion varies along different cutting directions on single-crystal work material, the primary crystallographic orientations that dictate the dislocation motion along activated slip systems is also considered [28]. Slip system activation is defined by the Schmid factor as follows [29]:

$$m = \cos\alpha \cos\beta \tag{11}$$

where $\alpha$ is the angle between the applied force vector and the slip plane normal, and $\beta$ is the angle between the applied force vector and the dislocation slip direction. Lastly, the directional influence of the magnetic field on slip activated dislocation motion in Eq. (10) is integrated with the Schmid factor as follows:

$$B_e^2 = B^2 \cos\theta \cos\lambda \cos\alpha \cos\beta \tag{12}$$

$$M = \cos\theta \cos\lambda \cos\alpha \cos\beta \tag{13}$$

where $M$ is defined as an orientation factor to account for the anisotropy relative to the magnetic field directions and the cutting directions. Therefore, the relative change of fracture toughness under the magnetic field is rewritten by substituting Eqs. (12) and (13) into Eq. (8) as follows:

$$\frac{K_{cB}}{K_{c0}} = \left(1 + \frac{k_\rho B_e^2}{B_0^2}\right)^{0.5} \left(\frac{c_0}{c_B}\right)^{1.5} = \left(1 + \frac{k_\rho B^2}{B_0^2} M\right)^{0.5} \left(\frac{c_0}{c_B}\right)^{1.5} \tag{14}$$

$$\frac{c_0}{c_B} \propto B^2 M \tag{15}$$

*2.5 Model application*

The orientation-dependent influences of the magneto-plastic effect are predicted during the micro-cutting of CaF$_2$ single crystal. In this work, four cutting orientations on the (111) plane are evaluated: $[\bar{1}\bar{1}2]$, $[\bar{1}10]$, $[11\bar{2}]$, and $[1\bar{1}0]$. Fig. 2 illustrates the relationship between the cutting directions and {100}⟨110⟩ primary slip system [30]. Four magnetic-field orientations (0°, 30°, 60°, and 90°) were assessed by varying the field angles relative to each cutting direction where 0° is the direction when the magnetic field direction and cutting direction are parallel to each other and 90° is where the magnetic field lines are perpendicular to the cutting direction.

The significance of the orientation-dependent $M$ values can be used to predict the ductile–brittle transition during micro-cutting with respect to the magnetic field directions and the cutting directions of the single crystal. Chen et al. [31] identified the correlation between the anisotropic fracture toughness and the ductile–brittle transition in micro-cutting of single-crystal CaF$_2$, which serves as the basis to employing the fracture toughness as a measure of predicting the ductile–brittle transition. Based on Eqs. (14) and (15), a higher $M$ value indicates an increase in fracture toughness, which would suggest that higher ductile–brittle transition can be expected with the appropriate selection of the magnetic field orientation relative to the cutting directions.

The values of $M$ with different cutting directions and magnetic field directions were calculated according to Eq. (13). As there are six slip systems in {100}⟨110⟩ family, six $M$ values for a specific cutting direction and magnetic field direction were calculated. Considering the multi-system slip characteristics in micro-cutting, the average values of $M$ were adopted to assess the relative change of ductile–brittle transition under magnetic field.

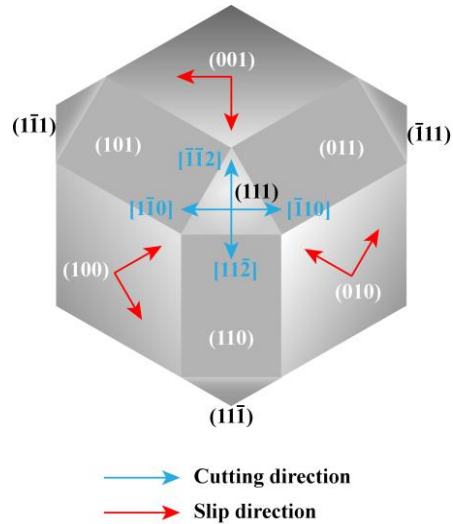

**Fig. 2.** Relationship between cutting directions and slip systems relative to (111) cutting plane.

Fig. 3(a) illustrates the calculated $M$ values versus six activated slip systems and the average values along the $[11\bar{2}]$ cutting direction at varying magnetic field directions. The magnitude of orientation factor $M$ reveals the impact of magnetic field on dislocation mobility during deformation. The largest $M$ value suggests a higher speed of dislocation motion in comparison to other slip systems, which occurs when the magnetic field was orientated along the 0° direction. However, the *M* in the slip system (001)[110] reduces to zero for the 90° magnetic field direction, which suggests that magnetic-excited dislocation slip will be restricted in such a configuration. The increase in dislocation mobility as described by the *M* value would suggest the reduction in dislocation density and thus relieve the stress concentration induced during dislocation pile-up for changes to the material fracture toughness and ductility. Golovin [9] asserted the influence of magnetic field on the initiation and propagation of cracks and revealed that the stepped-up dislocation motion in a magnetic field could improve the fracture resistance behaviour of materials by promoting stress relaxation in the regions of stress concentration.

The average *M* values reflect the effect of the interaction between the magnetic field and dislocation motion along activated slip systems on the fracture toughness at the specific cutting direction. The average M values correspond to the fracture toughness, which would serve as indicators of the ductile–brittle transition in micro-cutting. Fig. 3(b) shows the average *M* values along the four cutting directions. It suggests that the extent of improvement in ductile–brittle transition first reduces before increasing with larger magnetic-direction angles along [11$\bar{2}$], [1$\bar{1}$0] and [$\bar{1}\bar{1}$2] cutting directions. On the contrary, the ductile-brittle transition along the cutting direction [$\bar{1}$10] exhibit an opposite trend with varying magnetic field directions. Such a trend in ductile-brittle transition is a result of the anisotropic change in dislocation behaviour and fracture toughness with respect to magnetic field directions and cutting directions. The model firstly uncovers the orientation of magneto-plasticity in relation to magnetic field directions and crystallographic orientations.

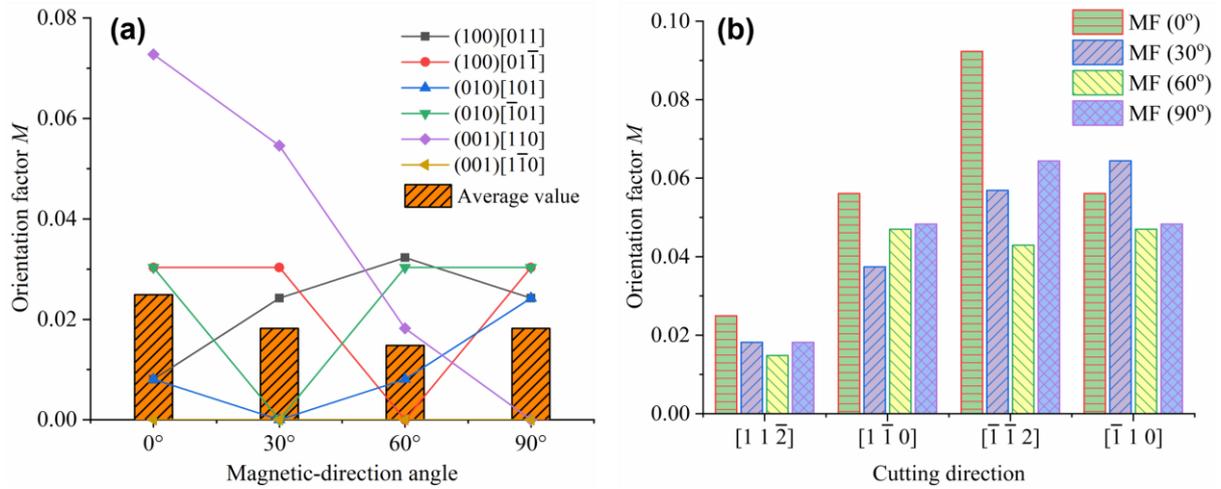

**Fig. 3.** (a) Calculated *M* values versus six activated slip systems and their averages value along [11$\bar{2}$] cutting direction at varying magnetic field directions; (b) Average values of *M* for all four cutting directions with respect to different magnetic field directions.

## 3 Experiments

*3.1 Work material*

Calcium fluoride (CaF$_2$) single-crystal with the (111) end plane and indexed edge orientations <110>/<211> was commercially procured as the work material for these experiments. The crystallographic orientation was indexed by X-ray diffraction (Bruker D8 Advance). The CaF$_2$ single crystal was a square workpiece with dimensions of 10×10×5mm. The material properties of CaF$_2$ are listed in Table 1. To investigate the influence of weak magnetic field on the ductile-mode cutting of $CaF_2$ single crystal, micro-indentation and micro plunge-cutting tests were carried out, respectively.

**Table 1.** Material properties of $CaF_2$.

| Crystal structure | Lattice constant | Density | Melting point | Hardness | Young's modulus |
|---|---|---|---|---|---|
| Cubic | a = 5.4626 Å | 3.18 g/cm$^3$ | 1360 °C | 158.3 Knoop | 75.8 GPa |

*3.2 Micro-indentation*

A Shimadzu HMV-2 Vickers micro-hardness test machine (Fig. 4(a)) was employed to characterize the material response under a weak magnetic field (20 mT) induced by a pair of permanent magnets. The intensity of the magnetic field was measured using a GM08-0493 Gaussmeter. The workpiece surfaces were double-side epi-polished on the (111) end plane to remove existing micro-cracks. Two different loads (0.25 N and 0.5 N) were applied in the micro-indentation tests with a loading rate of 0.5 mm/s and a dwell time of 15 s. Five repetitions were conducted for each test condition. Table 2 displays the testing parameters in micro-indentation. Residual impressions after micro-indentation were analyzed using an Olympus OLS5000 3D laser confocal microscope.

**Table 2.** Experimental parameters used in micro-indentation and micro plunge-cutting tests.

|  | Parameters | Values |
|---|---|---|
| Workpiece | Material | $CaF_2$ single crystal |
|  | Dimension | 10 × 10 × 5 mm |
|  | Crystal plane | (111) |
|  | Edge orientations | <110>/<211> |
| Micro-indentation tests | Load | 0.25 N, 0.5 N |
|  | Loading rate | 0.5 mm/s |
|  | Dwell time | 15 s |
|  | Magnetic field source | Permanent magnets |
|  | Magnetic field intensity | 20 mT |
| Micro plunge-cutting tests | Tool material | Single-crystal diamond (SCD) |
|  | Rake angle | -10° |
|  | Nose radius | 1.6 mm |
|  | Cutting speed $v_c$ | 20 mm/min |
|  | Undeformed chip thickness $d_c$ | 0–2 μm |
|  | Cutting length $L_c$ | 3 mm |
|  | Cutting directions | $[11\bar{2}]$, $[1\bar{1}0]$, $[\bar{1}\bar{1}2]$, $[\bar{1}10]$ |

| | |
|---|---|
| Magnetic field source | Electromagnet |
| Magnetic field intensity | 20 mT |
| Magnetic field directions | 0°, 30°, 60°, 90° relative to each cutting direction |

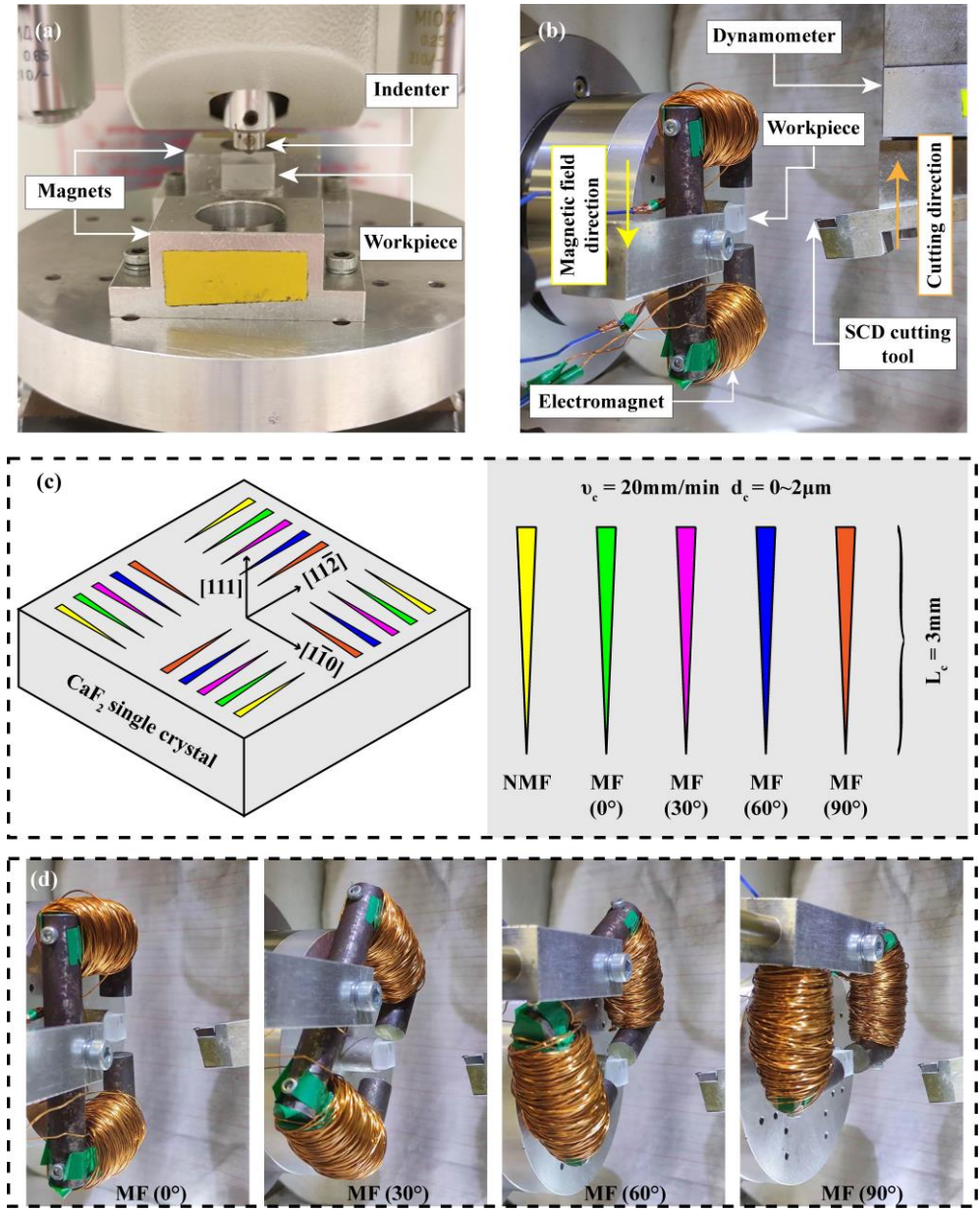

**Fig. 4.** (a) Experimental setup of micro-indentation tests under a weak magnetic field. (b) Magnetic field-assisted micro plunge-cutting setup on an ultra-precision machining centre. (c) Schematic of micro plunge-cutting tests. (d) Varying magnetic field direction during micro plunge-cutting tests. (NMF: no magnetic field. MF: magnetic field).

*3.3 Micro plunge-cutting*

Micro plunge-cutting experiments were performed on a Toshiba ULG-100 ultra-precision diamond turning machine as shown in Fig. 4(b). A single-crystal diamond (SCD) cutting tool with a -10° rake angle and nose radius of 1.6 mm was affixed to the UPM for trimming and plunge-cutting. A self-constructed electromagnet setup was positioned to generate a controlled magnetic field of 20 mT as shown in Fig. 4(b) and the schematic of the plunge-cutting tests is shown in Fig. 4(c). The undeformed chip thickness was gradually increased from 0 to 2 μm over a 3 mm cutting length with a constant cutting speed of 20 mm/min. Eddy current damping effects were neglected in view of the low cutting speed.

Micro-cutting was performed along the four different crystallographic directions on the (111) plane ($[11\bar{2}]$, $[1\bar{1}0]$, $[\bar{1}\bar{1}2]$, and $[\bar{1}10]$) by rotating the sample along the C-axis of the machine tool spindle. According to Smirnov [17], the magnetic driving effect on dislocation motion is not dependent on the polarity of the magnetic field. For each cutting direction, the magnetic field direction was adjusted at intervals of 30° ranging from 0° to 90° where 0° represents a magnetic field direction parallel to the cutting direction and 90° represents a magnetic field perpendicular to cutting direction as displayed in Fig. 4(d). Cutting forces (Fc) and thrust forces (Ft) were recorded using a Kistler 9256C1 dynamometer and type 5051A amplifiers during the micro-cutting process. After micro-cutting, the machined grooves were observed using the laser confocal microscope. The experimental parameters in micro plunge-cutting tests are shown in Table 2. The surface profile of the machined surfaces was measured using an Olympus LEXT OLS5500 laser confocal microscope. In addition, the machined subsurface was also evaluated by transmission electron microscopy (TEM) analysis (Talos 200FX FEI, USA). The cross-sectional TEM samples were prepared using the lift-out technique with a dual-beam focused ion beam (FIB) system (FEI HELIOS NanoLab 600i). TEM samples were selected in the ductile-mode regime.

**4 Results and discussion**

*4.1 Validation of material response under weak magnetic field by micro-indentation test*

Fig. 5 reveals height images of residual impressions after micro-indentation tests. The surface pile-up effect is a common occurrence during indentation tests where material around the indenter flows upwards above the initial material surface. Therefore, the pile-up height profiles along four different paths perpendicular edges of residual impressions in Fig. 5(b) were evaluated. Fig. 6 shows the surface pile-up height–distance curves of a pair of residual impressions in the absence and presence of the magnetic field, and under different loads of 0.25 N and 0.5 N. The measured pile-up height decreases with the application of the magnetic field as shown by the average measurements in Fig. 7. Under magnetic-free conditions, the average pile-up heights under indentation loads of 0.25 N and 0.5 N were 157 nm and 217 nm, respectively, while that of magnetic-assisted indentation dropped to 147 nm and 202 nm. Gale and Achuthan [32] claimed that the pile-up height is dependent on the statistically stored dislocation density $\rho_s$, where a higher density would result in a greater resistance to dislocation motion and promote pile-up formation. From Eqs. (2) and (4), a magnetic field would facilitate dislocation movement and lower the dislocation density to relieve the dislocation accumulation, thus reducing the pile-up height. These results affirm the notion of enhanced dislocation movement during deformation under the influence of a magnetic field.

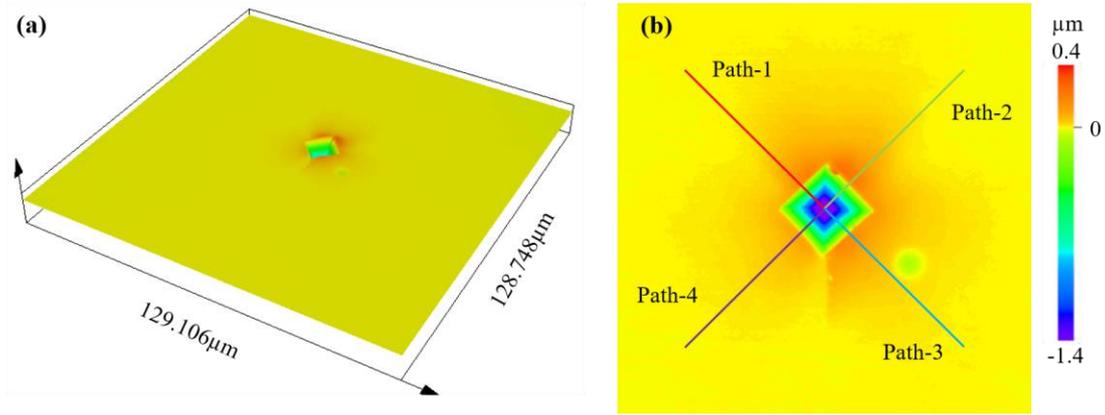

**Fig. 5.** Height images of residual impressions after micro-indentation tests. (a) 3D view; (b) 2D view.

As shown in Fig. 6, the pile-up height is substantially higher than the original sample surface and gradually reduces away from the indentation. Hence, the pile-up regions surrounding the indentation can be observed by the height values greater than zero in the top view images (Fig. 5). Fig. 8 displays a larger plastic zone with the application of the magnetic field during indentation. The plastic zone around the residual impressions reflects the plastic flow during indentation where its size serves as an indicator of the material yield strength [33]. The plastic zone size is inversely proportional to the material yield strength, where a lower yield strength would generate a larger plastic zone [34]. These results affirm the observations by Urusovskaya et al. [8] where the yield strength of different materials was found to decrease during deformation under a weak magnetic field.

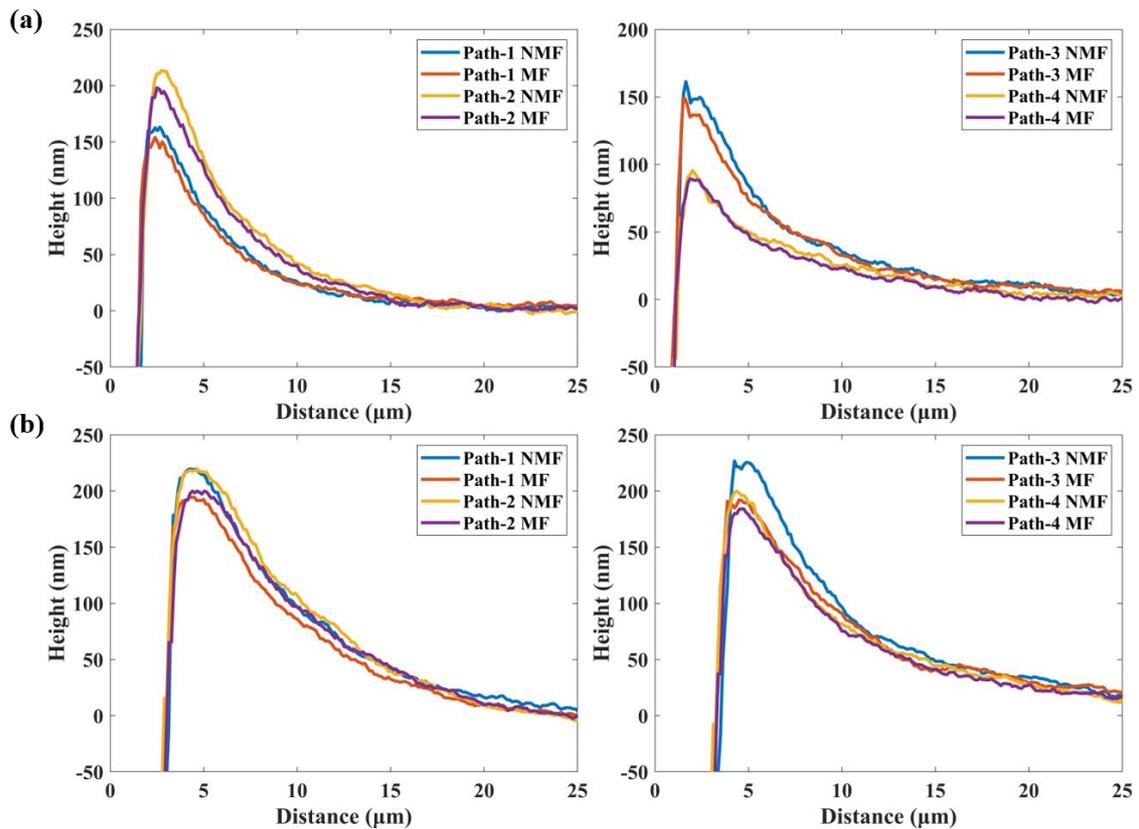

**Fig. 6.** Surface pile-up height–distance curves of a pair of residual impressions under magnetic field. (a) 0.25 N load; (b) 0.5 N load.

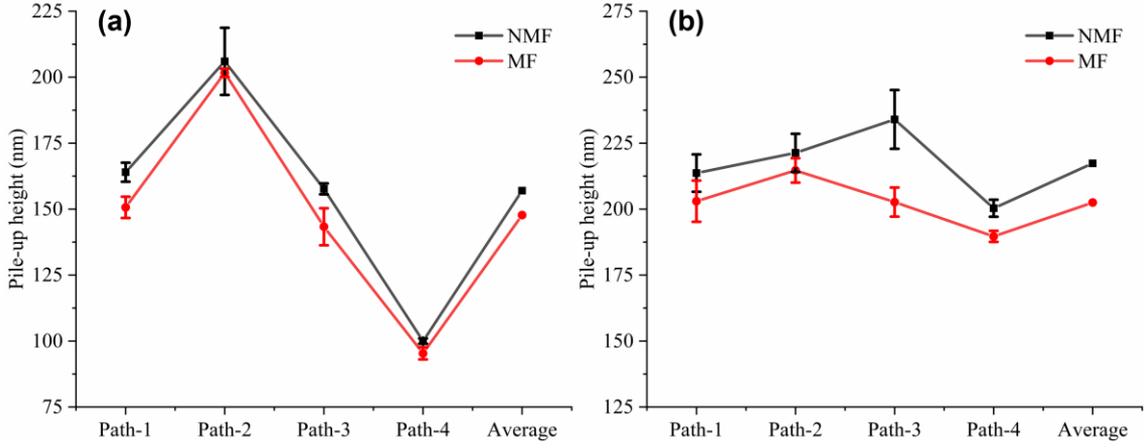

**Fig. 7.** Pile-up heights for each path and average pile-up height under magnetic field: (a) 0.25 N load; (b) 0.5 N load.

According to Busby et al. [35], the yield strength of the material $\sigma_y$ is assumed to be linearly correlated with the hardness $H$. Thus, Eq. (7) was rewritten by replacing the material hardness as a function of the yield strength as follows:

$$\frac{\sigma_{yB}}{\sigma_{y0}} = \frac{B_0^2}{B_0^2 + k_\rho B^2} \tag{16}$$

where $\sigma_{yB}$ and $\sigma_{y0}$ are the yield strength with and without a magnetic field. The size of plastic zone $r$ was determined by the yield strength $\sigma_y$ and the applied load $P$ [36]:

$$r = \sqrt{\frac{3}{2\pi} \frac{P}{\sigma_y}} \tag{17}$$

Following this, the magnetic-induced change of the plastic zone area under the same external load was found in Eq. (18):

$$\frac{r_B}{r_0} = \sqrt{\frac{B_0^2 + k_\rho B^2}{B_0^2}} = \sqrt{1 + \frac{k_\rho B^2}{B_0^2}} \tag{18}$$

where $r_B$ and $r_0$ are plastic zone size with and without a magnetic field. Eqs. (16) and (18) successfully demonstrate the weakening of the yield strength around residual impressions owing to the reduced dislocation density under the influence of the magnetic field, leading to the enlargement of the plastic zone.

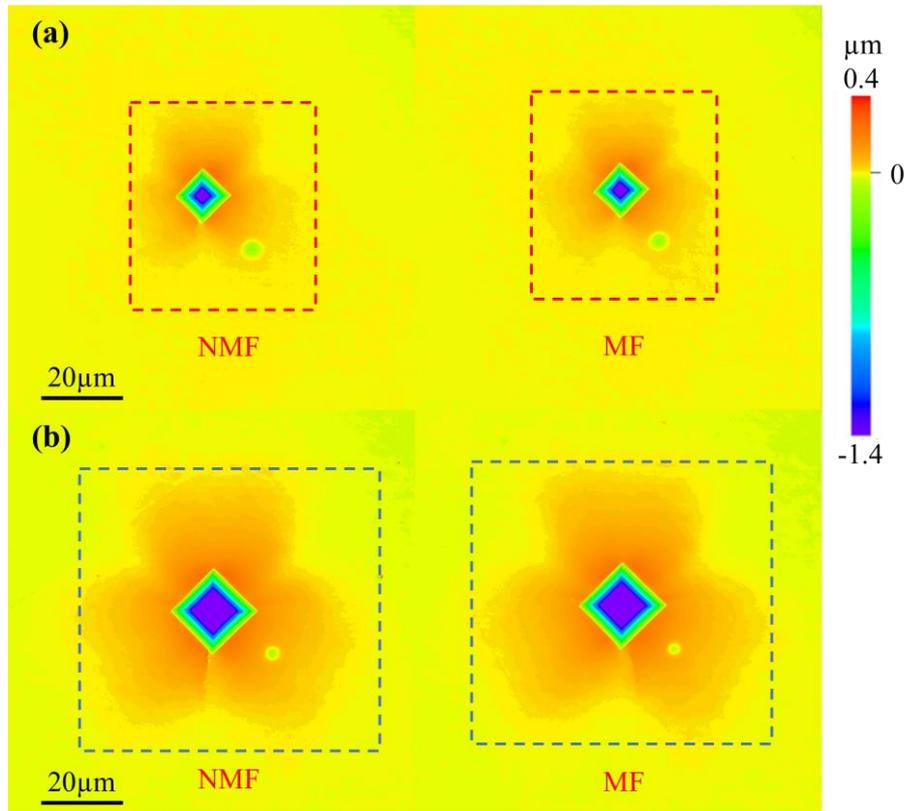

**Fig. 8.** Enlarged plastic zone with the application of the magnetic field: (a) 0.25 N load; (b) 0.5 N load.

Fig. 9 illustrates the optical residual impressions in the absence and presence of a magnetic field. For magnetic-free residual impressions, cracks appear around the indentations while such cracks were suppressed under the influence of the magnetic field. As mentioned in Section 2.2, crack initiation and propagation are strongly associated with the fracture toughness $K_C$. As shown by Eqs. (8) and (14), the larger fracture toughness enabled by the magnetic field will strengthen the ability to restrain crack propagation and result in a shorter crack length. It is evident that the material behaviour of brittle materials can be considerably modified under a weak magnetic field, which demonstrates the promising potential for a weak magnetic field to enhance the ductile-mode cutting of brittle materials. The next section offers a promising method to estimate the magnetic field-induced change in dislocation density and fracture toughness to quantify the magneto-plastic effect.

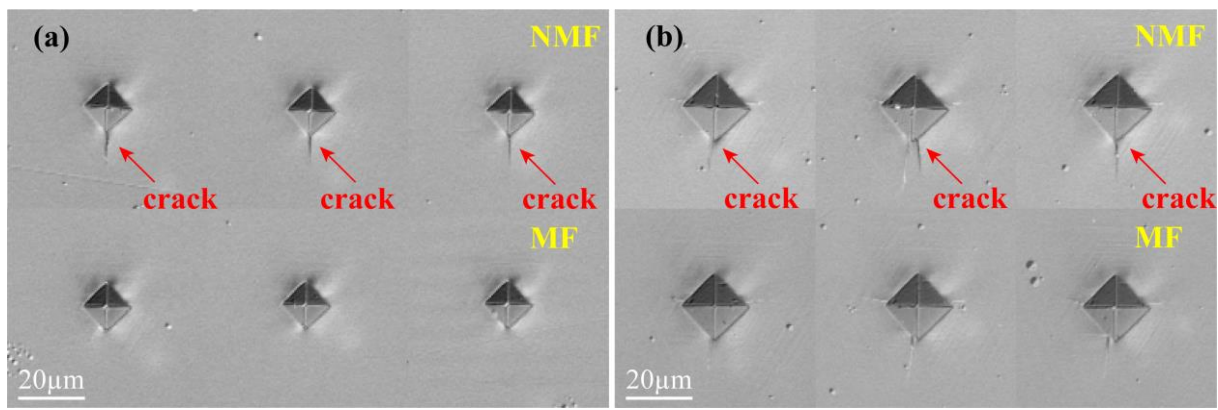

**Fig. 9.** Suppression of crack growth under magnetic field: (a) 0.25 N load; (b) 0.5 N load.

*4.2 Quantification of dislocation behaviour and fracture toughness induced by a weak magnetic field*

Dislocation slip traces induced during micro-indentation were also captured using the Olympus LEXT OLS5500 laser confocal microscope with a suitable image contrast setting as displayed in Figs. 10(a), 10(b) and S2. An obvious observation of dislocation slip traces is found along the $[1\bar{2}1]$ crystallographic orientation, which is oriented 60° from the main direction of crack that propagates along $[\bar{1}\bar{1}2]$. The sliding distance for the dislocation slip traces along $[1\bar{2}1]$ with the magnetic field (Figs. 10(b) and S2(b)) is much further compared to the distance measured without the magnetic field (Figs. 10(a) and S2(a)). The longer moving distance of dislocation slip traces under the magnetic field directly demonstrates the theory of magnetic field-accelerated dislocation mobility in Eq. (2), which is also consistent with the notion of a larger plastic zone size with the magnetic field in Fig. 8.

The magnetic field is understood to have an influence on the newly generated dislocation mobility and is independent of dislocation generation [5]. Hence, the number of emitted dislocations was assumed to be identical under the same loading conditions during indentation. As displayed in Figs. 10 (c) and (d), the line density of dislocation slip traces $\rho_L$ in Figs. 10(a), 10(b) and S2 was determined as follows:

$$\rho_L = \frac{N}{D} \tag{19}$$

where $N$ is the total number of activated dislocations and $D$ is the distance from the indentation centre to the utmost dislocation slip trace (Fig. 10(d)). The magnetic field-induced change in the line density of dislocation slip traces $\rho_L$ was then written as follows:

$$\frac{\rho_{LB}}{\rho_{L0}} = \frac{D_0}{D_B} \tag{20}$$

where $\rho_{tB}$ and $\rho_{t0}$ are the line density of dislocation slip traces in the presence and absence of a magnetic field, and $D_B$ and $D_0$ are the travelling distance of utmost dislocation slip trace with and without a magnetic field. The relative change in the line density of dislocation slip traces is synonymous with the alteration of total dislocation density [37] under the magnetic field. Thus, the density of slip traces was used as the quantitative estimation of magnetic field-induced alteration in the total dislocation density in Eq. (4). Subsequently, the change of material hardness and fracture toughness in a magnetic field was rewritten in Eqs. (6) and (8) as follows:

$$\frac{H_B}{H_0} = \frac{B_0^2}{B_0^2 + k_\rho B^2} = \frac{\sqrt{\rho_{tB}}}{\sqrt{\rho_{t0}}} \approx \frac{\sqrt{\rho_{LB}}}{\sqrt{\rho_{L0}}} = \frac{\sqrt{D_0}}{\sqrt{D_B}} \tag{21}$$

$$\frac{K_{cB}}{K_{c0}} = \left(\frac{H_0}{H_B}\right)^{0.5} \left(\frac{c_0}{c_B}\right)^{1.5} = \left(\frac{D_B}{D_0}\right)^{0.25} \left(\frac{c_0}{c_B}\right)^{1.5} \tag{22}$$

where $c_B$ and $c_0$ is appointed as the longest crack length with and without a magnetic field to facilitate the calculation.

Table S2 shows the travelling dislocation of utmost dislocation slip trace with and without a magnetic field, and the calculated results of magnetic field-induced change in dislocation density and fracture toughness. The dislocation density with the magnetic field was approximately 30% lower than that without the magnetic field. This corresponds to a 53.8% increase was observed in the fracture toughness when the magnetic field was applied. The quantitative results in dislocation density and fracture toughness adequately explain the observed material response during micro-indentation tests in correspondence to the theoretical model.

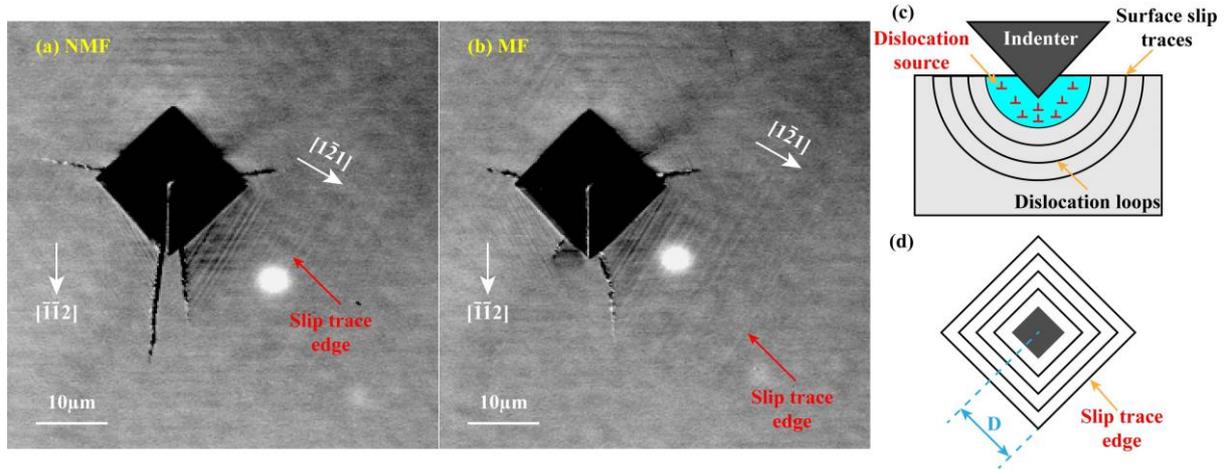

**Fig. 10.** Dislocation slip traces induced by the micro-indentation with a load of 0.5 N in the absence (a) and presence (b) of the magnetic field. Schematic of dislocation source, dislocation loops, surface slip traces during indentation: (c) Front view; (d) Top view [37].

*4.3 Magnetic field-induced ductile-brittle transition and its anisotropy in micro plunge-cutting*

    The change in fracture toughness under the influence of the magnetic field should be translatable to the ductile–brittle transition during micro-cutting of the single crystal. Fig. 11 depicts the machined surface morphology of micro-grooves along the $(111)[11\bar{2}]$ cutting direction. Plunge-cutting without a magnetic field resulted in distinct laminar cracks, represented by the blue regions in Fig. 11, that appear earlier on the machined surface at shallower cutting depths. An extension of the ductile-mode regime on the machined surface is observed with the application of the magnetic field. In addition, fewer cracks and smoother surface topography is observed in the brittle regime. The delayed ductile–brittle transition under the magnetic field also demonstrated the anisotropic feature where the length of the ductile-mode regime varied with magnetic-direction angles. The longest ductile-mode regime occurred when the magnetic field direction was perpendicular to the cutting direction (i.e., the 90° orientation).

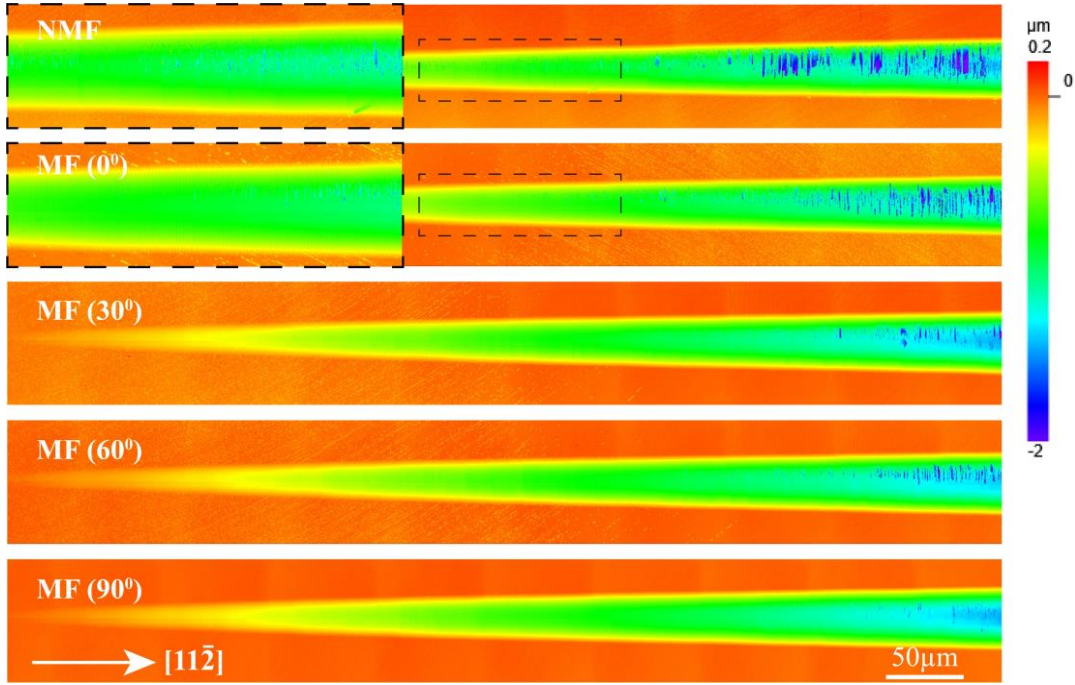

**Fig. 11.** Machined surface morphology of micro-grooves along [11$\bar{2}$] cutting direction with varying magnetic field directions.

Plots of the depth profiles for the plunge-cuts under different magnetic field conditions are shown in Fig. 12, where the critical undeformed chip thickness was determined. The ductile-mode regime was identified by the smooth profile of the plunge-cut, while the brittle-mode regime was determined by the erratic fluctuations in the depth profile owing to crack formation. The transition between the two types of depth profiles was defined as the critical undeformed chip thickness. The ductile–brittle transition without the magnetic field occurred at the lowest cutting depth of 512 nm in comparison to the critical values with the magnetic field, which ranged from 664 nm to 806 nm as a result of the magneto-plastic anisotropy.

Fig. 13 presents the critical undeformed chip thickness values during plunge-cut testing along the four main cutting directions and the combination of the different magnetic-direction angles. A general increase in the critical undeformed chip thickness can be observed across the four cutting directions when employing the external magnetic field. However, the trends in the enhanced ductile–brittle transition differed between each crystallographic orientation when employing different magnetic field orientations. The critical undeformed chip thickness increased progressively with the increase in magnetic-direction angles for the cut along [11$\bar{2}$]. However, the trend of the ductile-mode cutting regime was not as progressive during cut along the [$\bar{1}\bar{1}$2] direction. The ductile–brittle transition first decreased (0–30°) before increasing again at higher magnetic-direction angles (30–90°). The [1$\bar{1}$0] cutting direction showed a higher critical undeformed chip thickness values with the 0° and 60° magnetic-direction angles. An opposite occurred for the [$\bar{1}$10] cutting direction where lower values were recorded for the 0° and 60° magnetic-direction angles.

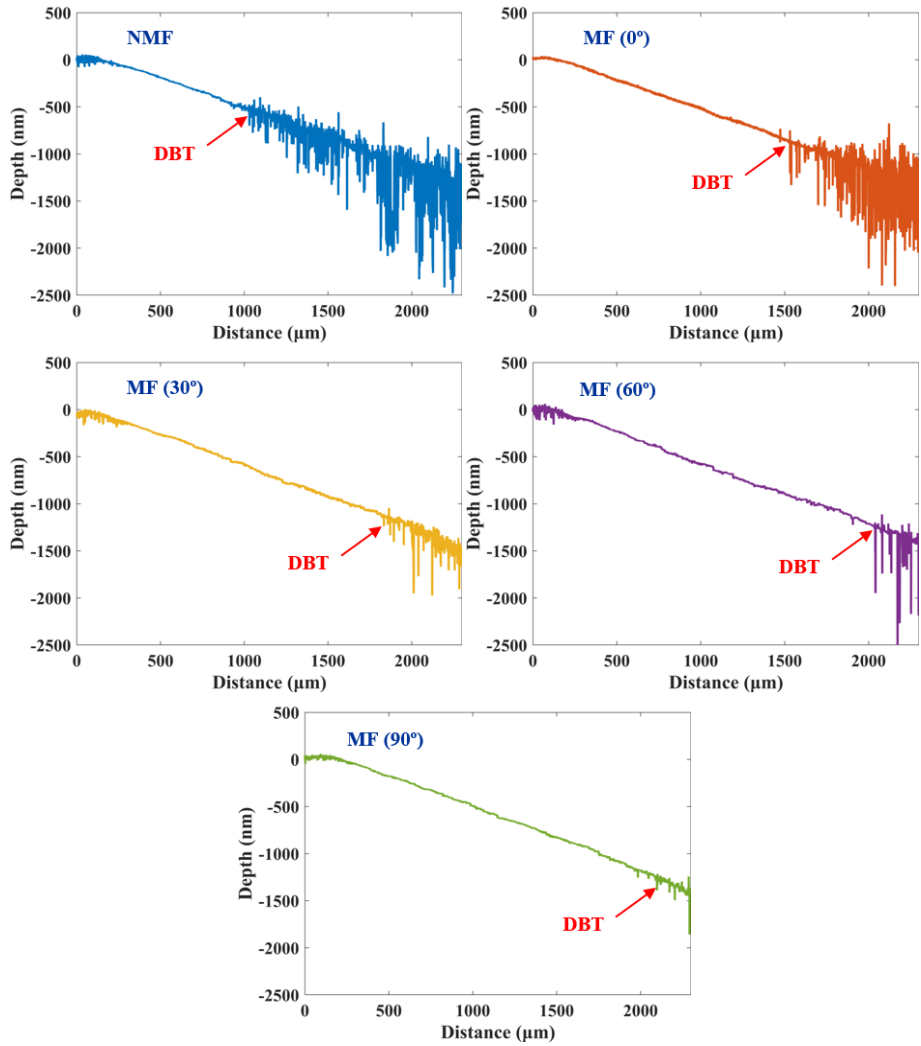

**Fig. 12.** Plunge-cutting depth-distance curves of micro-grooves along $[11\bar{2}]$ cutting direction with varying magnetic field directions. DBT: ductile-to-brittle transition.

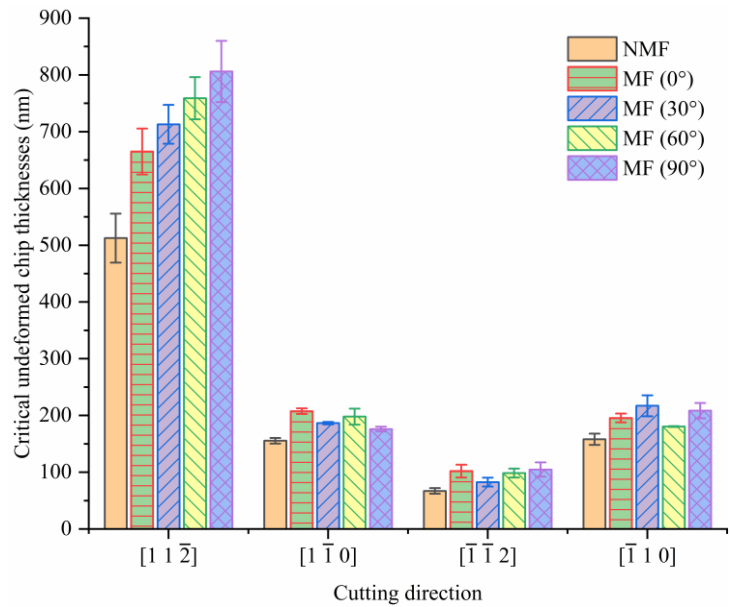

**Fig. 13.** Critical undeformed chip thickness at ductile–brittle transition position with varying magnetic field directions for four cutting directions.

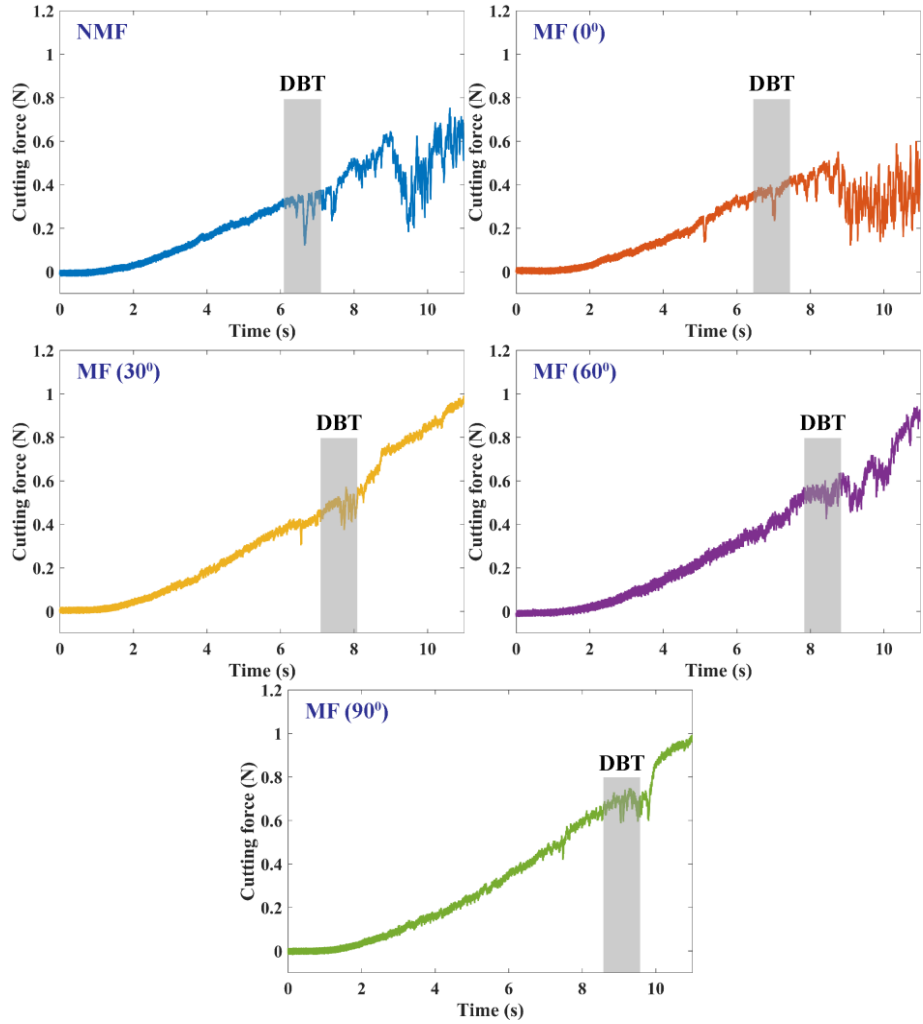

**Fig. 14.** Cutting force curves along $[11\bar{2}]$ cutting direction at varying magnetic field directions.

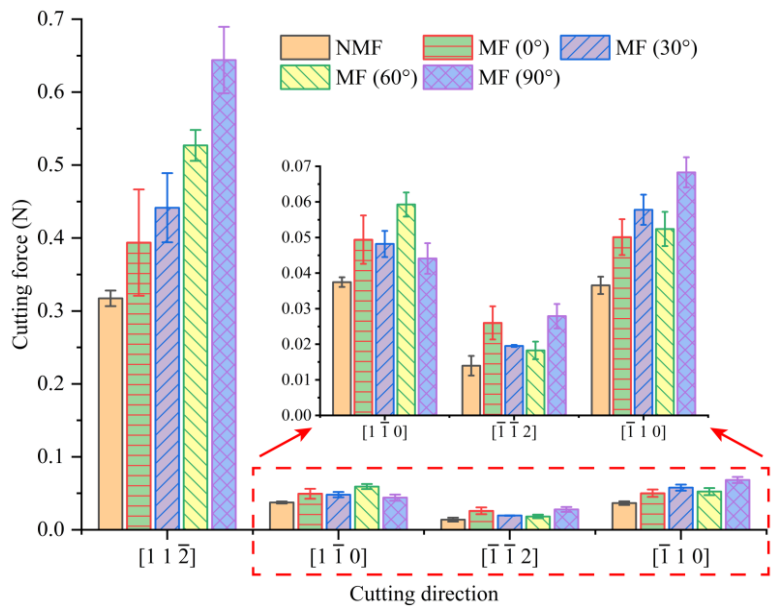

**Fig. 15.** Cutting force at ductile–brittle transition position with varying magnetic field directions for four cutting directions.

As illustrated in Fig. 14, the cutting force with magnetic-assisted plunge-cutting along [11$\bar{2}$] remained similar despite the delay in brittle-mode cutting. Ductile-mode cutting was identified by the stable force progression, which transits into the brittle-mode at the later stage where the force profiles showed instability with distinct fluctuations. Alike the orientation-dependent improvement in critical undeformed chip thickness, the cutting force plots showed delays in the ductile–brittle transition with increasing magnetic-direction angles when cutting along [11$\bar{2}$]. The change in cutting forces at the ductile–brittle transition under varying magnetic field directions along other three cutting directions are exhibited in Fig. 15, which is consistent with the directionally-dependent change in the critical undeformed chip thickness, relative to the magnetic field and cutting directions, as described in Fig. 13. These results in critical undeformed chip thickness and cutting force at the ductile–brittle transition demonstrate the orientation-dependency of the magnetic field direction relative to the crystallography.

The magnetic-induced changes in the critical undeformed chip thickness and cutting forces at the ductile–brittle transition position were identified and the ratio between the magnetic-assisted cutting results and the conventional cutting results was defined as the relative change (i.e., magnetic-field/no-magnetic-field). Figs. 16(a) and (b) illustrate the relative change in the critical undeformed chip thickness and cutting force at the ductile–brittle transition during plunge-cutting relative to varying magnetic field and cutting directions. From Eq. (15) and Fig. 3(b), the relative improvement of ductile–brittle transition in micro-cutting shows a strong correlation with the theoretical orientation factor $M$. The difference along the [11$\bar{2}$] cutting direction may be a result of the single-crystal lattice rotation induced by shear deformation or rotational deformation during the micro-cutting process of $CaF_2$ single crystals [38,39]. According to Zhang et al. [40], the lattice misorientation induced by crystal lattice rotation is heavily dependent on the crystal obstacle density and the resistance to dislocation motion during deformation and free-standing on the crystallographic orientation. In this work, the cutting force plays a critical role in the generation of obstacles and a higher force results in the formation of more obstacles that resist dislocation movement. As shown in Fig. 15, the cutting forces along [11$\bar{2}$] are approximately 10 times larger than other cutting directions, which suggests that a higher degree of lattice misorientation is likely to appear for the [11$\bar{2}$] cutting direction.

As displayed in Fig. 17, the cutting process results in a lattice rotation angle $\psi$ between the original crystal and the rotated crystal. The rotation of crystalline lattice triggers a change in the slip directions and slip planes relative to the original reference frame, which significantly affects the values of $M$ in Eq. (13) due to its high dependence with the coordinates of slip systems relative to the cutting direction and magnetic field direction. Therefore, the orientation factor in Eq. (13) was corrected considering the crystal rotation as follows:

$$M = cos\,\theta'\,cos\,\lambda'\,cos\,\alpha'\,cos\,\beta' \qquad (23)$$

where $\theta'$ is the angle between magnetic field direction and rotated slip plane normal, $\lambda'$ is the angle between magnetic field direction and rotated slip direction, $\alpha'$ is the angle between cutting direction and rotated slip plane normal, and $\beta'$ the angle between cutting direction and rotated slip direction. To assess the influence of the rotated crystalline lattice, a lattice rotation angle of 5° and 10° was assumed in Eq. (23) to respectively determine the updated values of $M$ as plotted in Figs. 16(c) and (d). An improved correspondence is found between the relative change in the critical undeformed chip thickness and cutting force at the ductile–brittle transition, and values of $M$ with a lattice rotation angle of 5° and 10°. As compared to other cutting directions that exhibit consistency between the experimental results and the theoretical $M$ values with the 5° lattice rotation angle, the [11$\bar{2}$] cutting direction shows a better agreement with the 10° lattice rotation angle, which affirms the notion that a higher degree of lattice misorientation is likely to occur along the $[11\bar{2}]$ cutting direction due to the force-induced obstacle density on the subsurface.

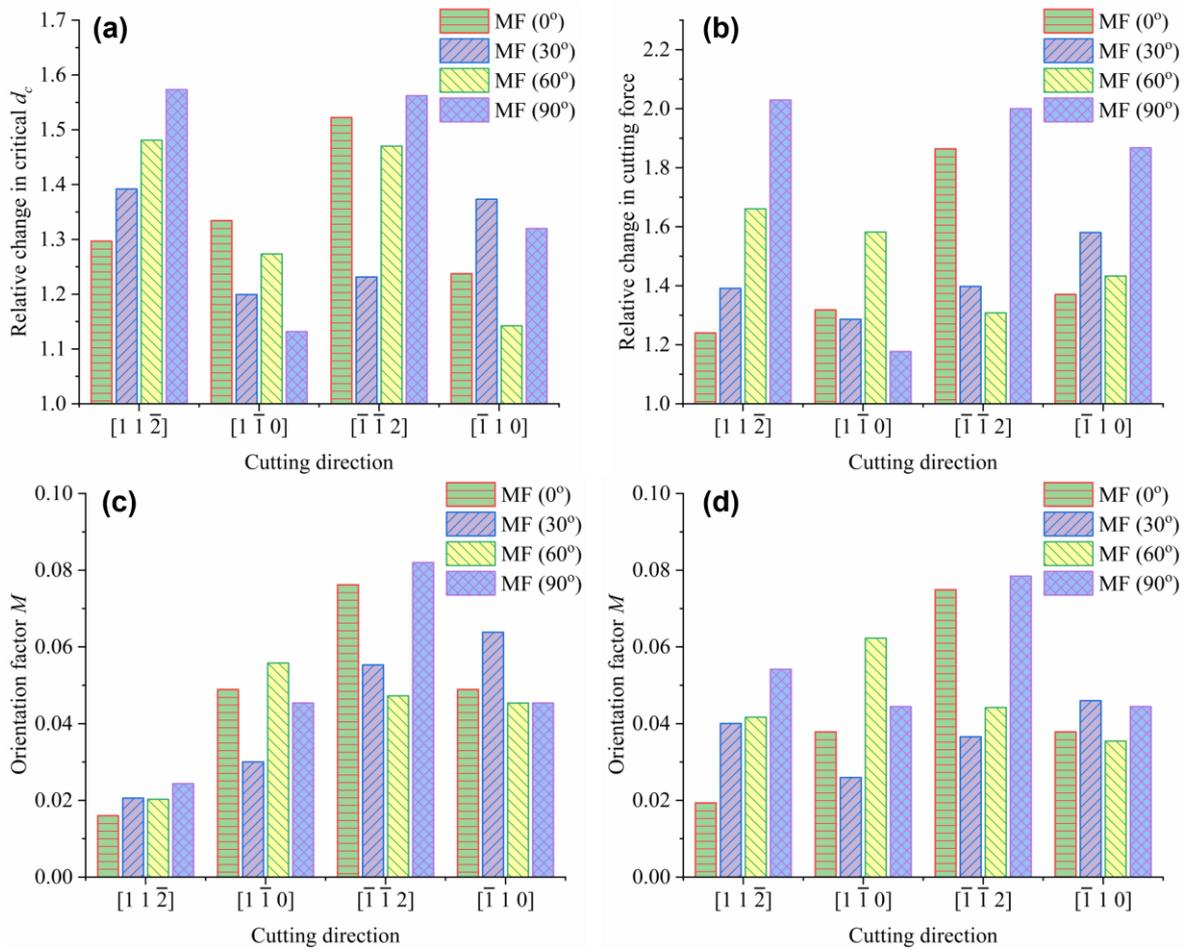

**Fig. 16.** (a) Relative change in critical undeformed chip thickness; (b) Relative change in cutting force at ductile–brittle transition; (c) Value of $M$ at 5° lattice rotation angle; (d) Value of $M$ at 10° lattice rotation angle with different cutting directions under varying magnetic field directions. $d_c$: undeformed chip thickness.

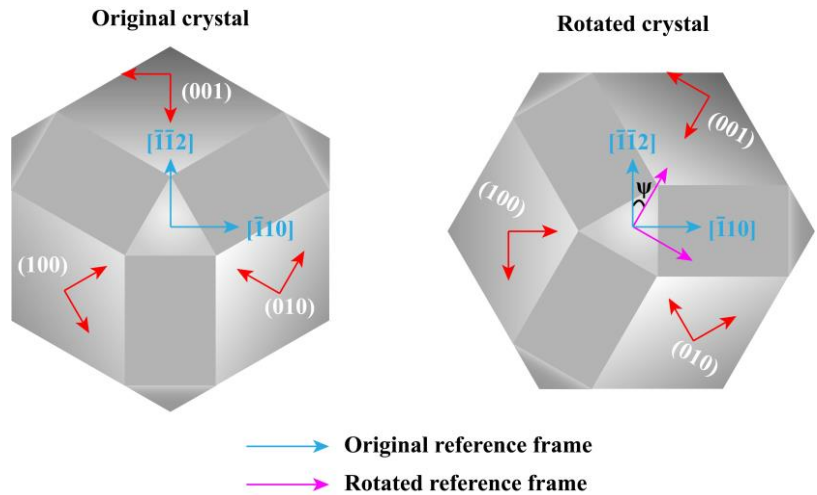

**Fig. 17.** Schematic of crystal rotation induced by shear deformation or rotational deformation during the cutting process.

*4.4 Subsurface microstructure analysis by TEM*

    A transmission electron microscope (TEM) was employed to observe the effect of magnetic field on the machined-induced subsurface damage and the change of crystal structure. Figs. 18(a), 19(a), and 20(a) provide a view of the cross-sectional TEM (XTEM) subsurface structure for $CaF_2$. These TEM images correspond to the subsurface of the plunge-cuts at a cutting depth of 500 nm along the $[11\bar{2}]$ cutting direction. From the bottom to the machined surface, the single-crystal layer (region 1), transition zone (region 2) and deformed layer (region 3) are observed in the figures. The subsurface of the magnetic-free cut reveals an obvious longitudinal crack that is aligned along the $(11\bar{1})$ cleavage plane at approximately 70.5° from the surface and a subsurface damaged layer that is 122 nm thick (Fig. 18(a)), which is a result of the joint occurrence of ductile-mode and brittle-mode cutting. In comparison, a crack-free subsurface is shown with the application of the magnetic field as displayed in Figs. 19(a) and Fig. 20(a). In accordance with the suppression of crack growth under the magnetic field in Fig. 9 and Eq. (14), the increased fracture toughness must be responsible for the enhanced crack-free subsurface after applying the magnetic field.

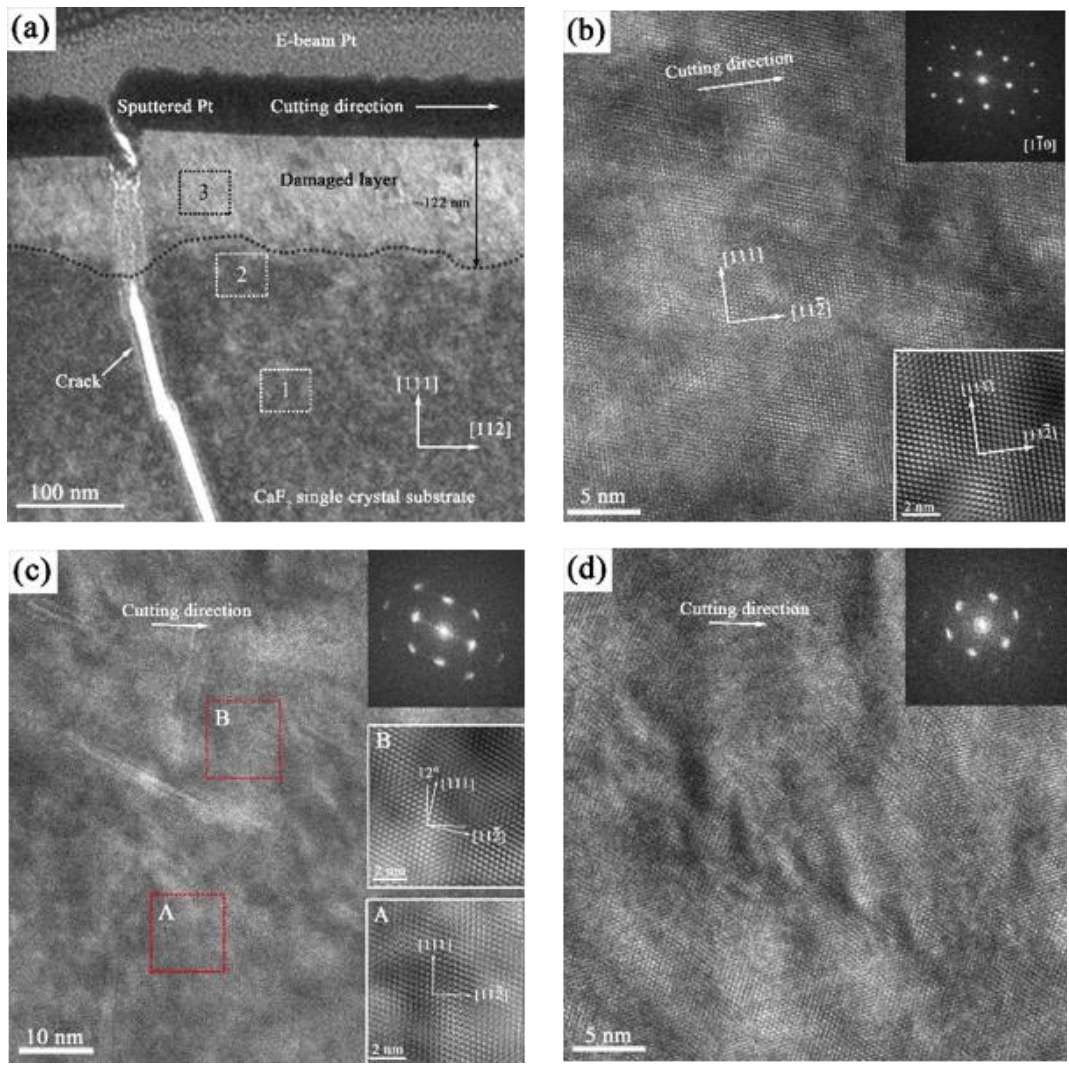

**Fig. 18.** Machined subsurface of $CaF_2$ single crystal along $[11\bar{2}]$ cutting direction without magnetic field: (a) cross-sectional transmission electron microscopy TEM (XTEM) overview of subsurface damage; (b), (c), (d) high-resolution TEM (HRTEM) images and selected area electron diffraction (SAED) patterns of regions 1, 2, 3 in (a).

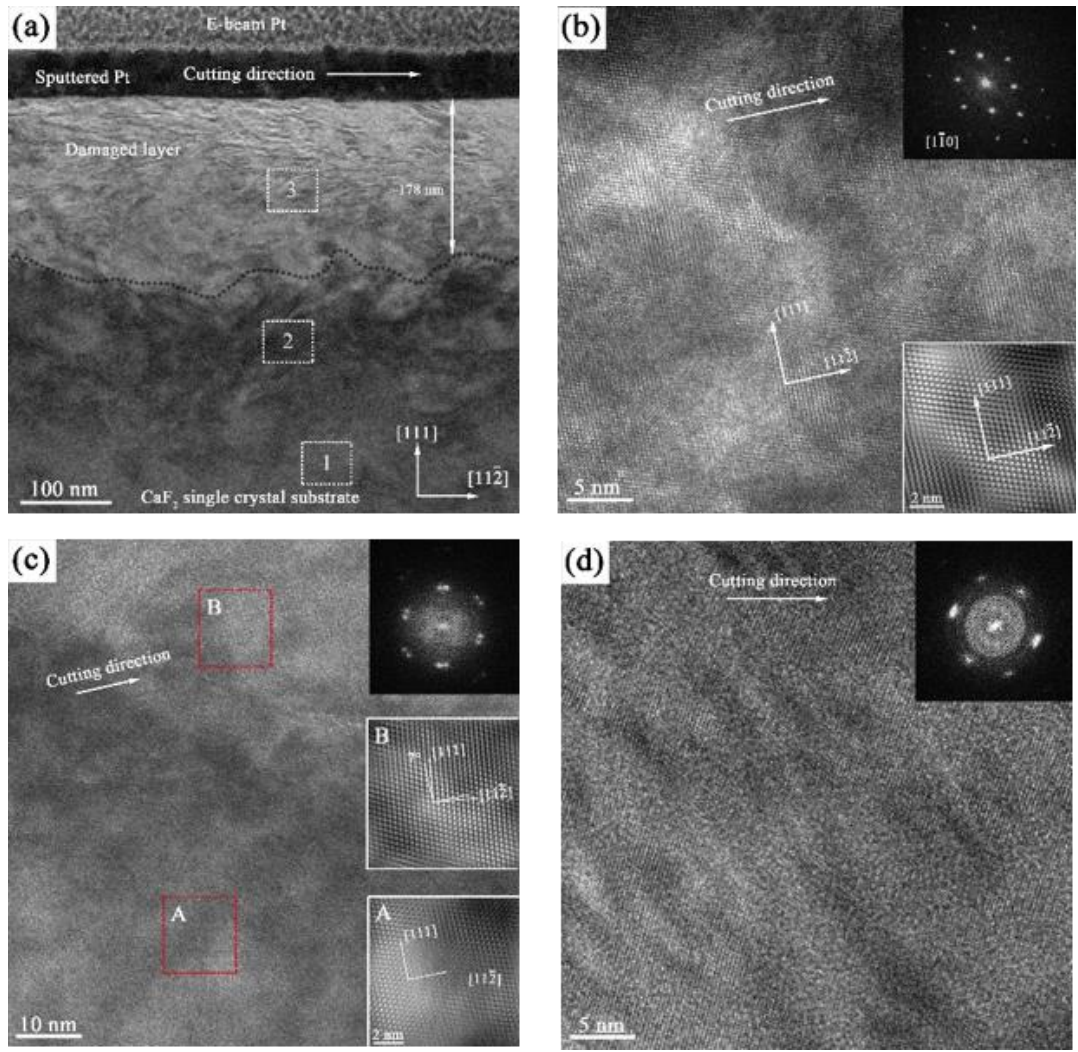

**Fig. 19.** Machined subsurface of CaF$_2$ single crystal along [11$\bar{2}$] cutting direction with a magnetic field (0°). (a) cross-sectional transmission electron microscopy TEM (XTEM) overview of subsurface damage; (b), (c), (d) high-resolution TEM (HRTEM) images and selected area electron diffraction (SAED) patterns of regions 1, 2, 3 in (a).

    The machined subsurface structure also exhibits anisotropic traits of the magneto-plastic effect under varying magnetic field orientations. A shallower subsurface damaged layer with a depth of 134 nm (Fig. 20(a)) appeared for 90° magnetic-direction angle in comparison with that of 178 nm (Fig. 19(a)) at 0° magnetic-direction angle. These subsurface damaged layers are thicker than the machined subsurface without the magnetic field, which is attributed to the release of the elastic strain energy induced during cutting that was dissipated into surface area generation during brittle fracture. In Figs. 19(d) and 20(d), selected area electron diffraction (SAED) patterns were employed to reveal the crystallinity of the deformed layer at 0° and 90° magnetic-direction angles. The deformed layer possessed a better crystallinity at 90° magnetic-direction angle. Similar observations are found in the transition zone of 0° (Fig. 19(c)) and 90° (Fig. 20(c)) magnetic-direction angles. From the SAED patterns of the transition zone in Figs. 18(c), 20(c), and 21(c), a lattice rotation angle of approximately 10° is observed, which further affirms the coherency between the theoretical values of $M$ and the relative change of critical undeformed chip thickness and cutting force at ductile-brittle transition when compared with other assumed lattice rotation

angles along $[11\bar{2}]$ cutting direction in Fig. 16. In addition to predicting the relative change in the critical undeformed chip thickness, the orientation factor $M$ is also able to estimate the stress level on the subsurface. Owing to a higher value of $M$ at 90° magnetic-direction angle along the $[11\bar{2}]$ cutting direction (Fig. 20(d)), there will be a lower stress level on the subsurface due to accelerated dislocation movement and reduced dislocation density based on the Eqs. (2), (4), (12) and (13), which results in weaker subsurface damage and better crystallinity relative to 0° magnetic-direction angle. This further demonstrates the orientation-dependency of the magneto-plastic effect under varying magnetic field directions during micro-cutting single-crystals.

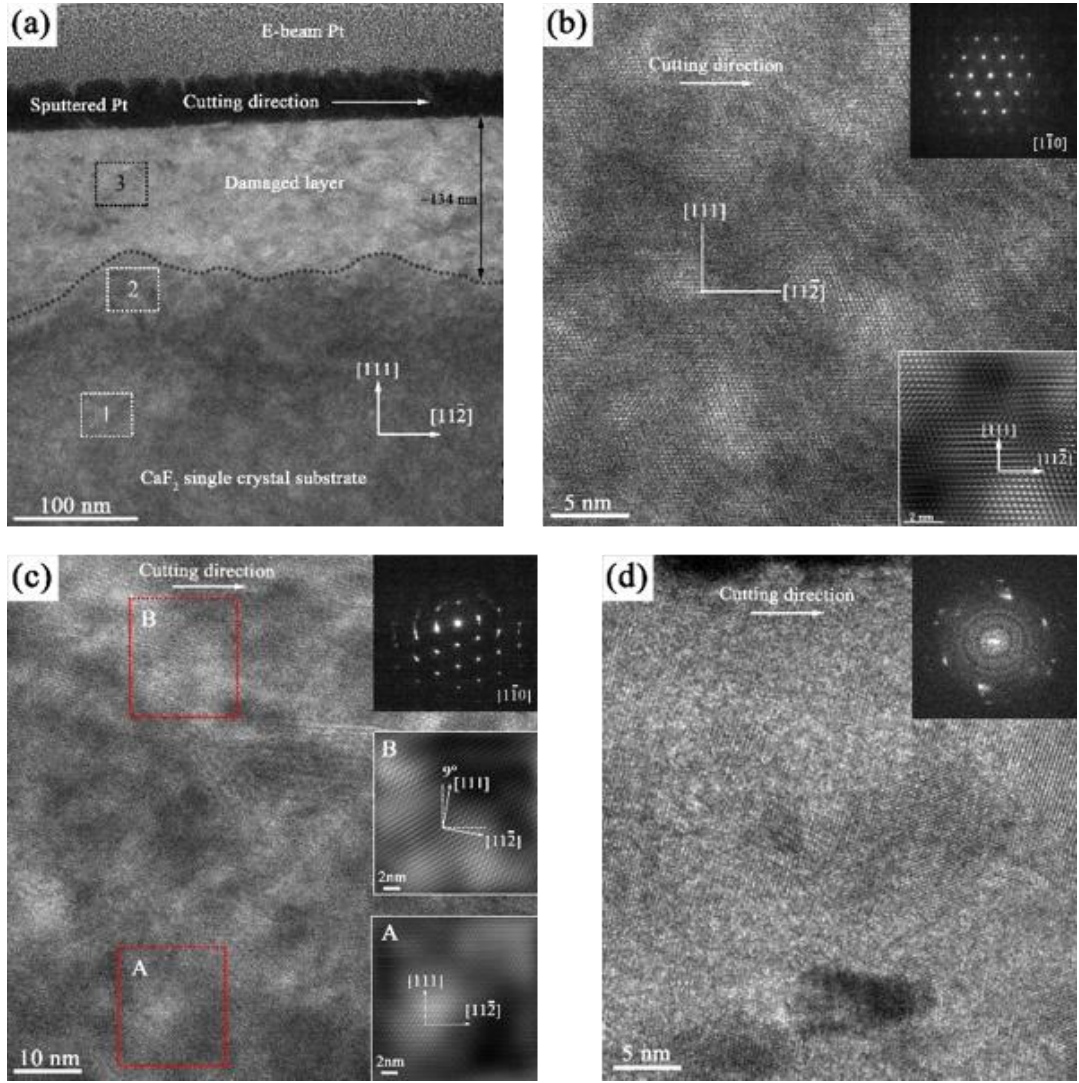

**Fig. 20.** Machined subsurface of $CaF_2$ single crystal along $[11\bar{2}]$ cutting direction with a magnetic field (90°). (a) cross-sectional transmission electron microscopy TEM (XTEM) overview of subsurface damage; (b), (c), (d) high-resolution TEM (HRTEM) images and selected area electron diffraction (SAED) patterns of regions 1, 2, 3 in (a).

## 5 Conclusions

The magneto-plastic effect was studied in the micro-cutting of single-crystal $CaF_2$ under a weak magnetic field. This encompassed substantiated reasoning for the material plasticity to be affected by an externally applied magnetic field through density functional theory calculations, which laid the groundwork for further development of a qualitative analytical model to derive the influence of a magnetic field on the mechanical properties of the material. Micro-indentation tests and ultra-precision micro-cutting tests were performed to evaluate the magneto-plastic effect on $CaF_2$ during deformation. The underlying mechanism for magnetic field-induced ductile–brittle transition in micro-indentation and micro-cutting was explained considering anisotropic dislocation behaviour and fracture toughness based on the coupled effect of magneto-plasticity and crystal plasticity. The anisotropic magneto-plastic model firstly uncovers the orientation of magneto-plasticity in relation to magnetic field direction and crystallographic orientation. The major conclusions of this study are summarized as follows:

(1) Enhanced plasticity is observed after applying the magnetic field, which includes the weakened surface pile-up effect, enlarged plastic zone, and suppressed crack propagation around residual impressions in micro-indentation tests.

(2) Density functional theory calculations indicate the possibility of non-singlet state defects, which can interact with the magnetic field and potentially excite low-energy triplet states at dislocations to activate the dislocation depinning process.

(3) Magnetic field-induced changes in dislocation density and fracture toughness are quantified by visual observation of dislocation slip traces in micro-indentation, which shows a 30% reduction in dislocation density and a 53.8% increase in fracture toughness.

(4) Implementation of the weak magnetic field during micro-cutting greatly enhances ductile-mode cutting with an increase in the ductile–brittle transition along the cutting direction $[11\bar{2}]$, manifested by the critical undeformed chip thickness from 512 nm and cutting force from 0.317 N under conventional cutting conditions to a range of 664–806 nm and 0.394-0.644 N with magnetic assistance, respectively.

(5) The improvement in ductile–brittle transition exhibits an orientation-dependent characteristic relative to both the magnetic field directions and the cutting directions. The ductile–brittle transition increases with rising magnetic-direction angles along the cutting direction $[11\bar{2}]$, while the critical values along the $[\bar{1}\bar{1}2]$ cutting direction first decreases before increasing again with the magnetic-direction angles. The ductile-mode cutting regime along the $[1\bar{1}0]$ cutting direction expresses an opposite magnetic-anisotropic improvement where the thickness increases before decreasing.

(6) The subsurface microstructure also shows the magnetic-anisotropy characteristic along the cutting direction $[11\bar{2}]$. A smaller subsurface damage layer and a better crystallinity occur when the magnetic field is oriented vertically to the cutting direction.

**Declaration of Competing Interest**

We declare that we have no conflict of interest.

**Acknowledgements**

This work is supported by the Singapore Ministry of Education Academic Research Funds (T2EP50120-0021).

# Supporting Information

# Effect of a weak magnetic field on ductile–brittle transition in micro-cutting of single-crystal calcium fluoride

Yunfa Guo[1], Yan Jin Lee[1], Yu Zhang[1], Anastassia Sorkin[1], Sergei Manzhos[2,*] and Hao Wang[1,*]

[1] *Department of Mechanical Engineering, National University of Singapore, 9 Engineering Drive 1, Singapore 117575, Singapore*

[2] *School of Materials and Chemical Technology, Tokyo Institute of Technology, Ookayama 2-12-1, Meguro-ku, Tokyo 152-8552, Japan*

* Corresponding authors. Emails: mpewhao@nus.edu.sg (H. Wang); manzhos.s.aa@m.titech.ac.jp (S. Manzhos)

## 1 DFT calculations

Spin-polarized density functional theory (DFT) calculations [1,2] were performed in SIESTA. [3,4]. The PBE exchange-correlation functional [5] was used and a DZP (double-ζ polarized) basis set. The density cutoff was 200 Ry. The density matrix convergence criterion was $1 \times 10^{-5}$ and the atomic forces and pressure were converged to 0.005 eV/Å and 0.005 GPa, respectively. During the calculations we used Troullier-Martins pseudopotentials [6]: $2s^22p^5$ electrons were treated as valence electrons for F and $4s^2$ for Ca. Most calculations were performed with a cubic simulation cell consisting of 32 $CaF_2$ units (before the introduction of the defect). The Brillouin zone was sampled with 3×3×3 $k$-points. For the stacking fault, the initial sample was twice bigger than the cubic one used for previous calculations with 3×3×1 $k$-points. The size of the sample used for dislocations was 22.00×5.5×33.00 Å (before optimization) or 96 $CaF_2$ units, the two dislocations and the stacking fault between them were created by cutting out 7 $CaF_2$ units in a row along the z-axis. The $k$-point sampling used for this case was 1×3×1.

Defect formation energies were computed as

$$E_f = \frac{E_d - (E_{ideal} + nE_{Ca} + mE_F)}{n+m} \quad (1)$$

where $E_d$ is the energy of the simulation cell with defects, $E_{ideal}$ is the energy of the corresponding simulation cell of ideal $CaF_2$, $n$ is the number of inserted or removed Ca atoms, $m$ is the number of inserted or removed F atoms. As reference energies (taken with a negative sign for removed atoms), $E_{Ca}$ and $E_f$ are taken as the energy of one Ca atom in face centred cubic (FCC) Ca and one half of the energy of an $F_2$ molecule (places in a box of size 10×10×10 Å), respectively. While these reference states are different from the chemical potentials of Ca and F in $CaF_2$, they do allow relative comparisons of defect formation energies among defect types.

The table below lists formation energies, spin states and, where applicable (for systems with singlet ground state), singlet-triplet gaps. Moreover, the perfect $CaF_2$ crystal is drawn in Fig. S1(a), while the defects used in this work are drawn in Figs. S1(b1)-(k).

**Table S1.** Lowest energy state, formation energies, and singlet-triplet gaps for various defects in $CaF_2$.

| Defect | Lowest energy state | Formation energy (eV) | Singlet-triplet gap (eV) |
|---|---|---|---|
| Perfect $CaF_2$ (Fig. S1(a)) | singlet | 0.00 | 6.316 |
| Vacancy Ca (Fig. S1(b1)) | triplet | 13.14 | |
| Vacancy F (Fig. S1(b2)) | doublet | 7.39 | |
| 2 vacancies of Ca (Fig. S1(c1)) | quintet | 26.18 | |
| 2 vacancies of Ca (Fig. S1(c2)) | quintet | 26.03 | |
| 2 vacancies of F (Fig. S1(d1)) | singlet | 13.57 | 1.343 |
| 2 vacancies of F (Fig. S1(d2)) | triplet | 14.79 | |
| 1 vacancy Ca and 1 vacancy F (Fig. S1(e1)) | doublet | 14.80 | |
| 1 vacancy Ca and 1 vacancy F (Fig. S1(e2)) | doublet | 15.81 | |
| 1 vacancy Ca and 2 vacancies F (Fig. S1(f1)) | singlet | 16.83 | 4.700 |
| 1 vacancy Ca and 2 vacancies F (Fig. S1(f2)) | singlet | 17.05 | 4.149 |
| 1 vacancy Ca and 2 vacancies F (Fig. S1(f3)) | singlet | 17.78 | 4.412 |
| 1 vacancy Ca and 2 vacancies F (Fig. S1(f4)) | singlet | 19.67 | 3.879 |
| 2 vacancies Ca and 4 vacancies F (Fig. S1(g1)) | singlet | 33.13 | 4.813 |
| 2 vacancies Ca and 4 vacancies F (Fig. S1(g2)) | singlet | 31.11 | 4.895 |
| Frenkel defect (Ca) (Fig. S1(h1)) | doublet | 8.99 | |
| Anti-Frenkel defect (F) (Fig. S1(h2)) | singlet | 10.28 | 6.69 |
| Ca interstitial (Fig. S1(i1)) | singlet | 12.90 | 2.763 |
| F interstitial (Fig. S1(i2)) | doublet | 0.46 | |
| 2 stacking faults (2D) (Fig. S1(j)) | singlet | 0.052* | 0.044* |
| 2 Dislocations with a stacking fault between them (1D) (Fig. S1(k)) | singlet | 0.090* | 0.064* |

* The energies per atom

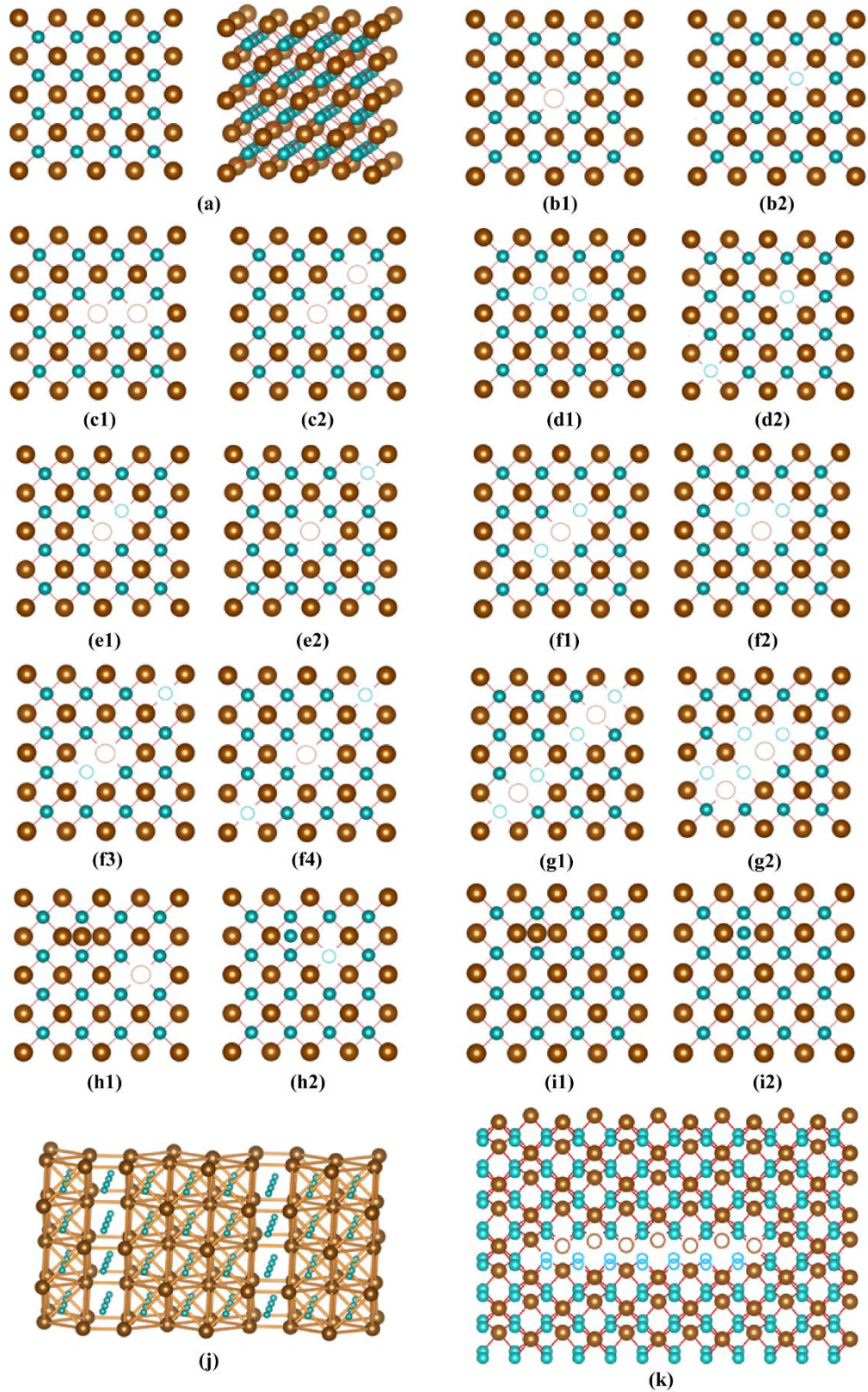

**Fig. S1.** Potential intrinsic lattice defects in CaF$_2$ single crystal used in this work. (a) The perfect crystal of CaF$_2$: front view and side view (Brown balls are Ca atoms, blue balls are F atoms). Simple vacancies in CaF$_2$: (b1) vacancy in Ca site, (b2) vacancy in F site. (c1), (c2) Two configurations of two vacancies in Ca sites. (d1), (d2) Two configurations of two vacancies in F sites. (e1), (e2) Two configurations of one vacancy in Ca site and one vacancy in F site. (f1)-(f4) Four configurations of one vacancy in Ca site and two vacancies in F sites. (g1), (g2)

Two configurations of two vacancies in Ca sites and four vacancies in F sites. Frenkel (h1) and anti-Frenkel (h2) defects. Ca (i1) and F (i2) interstitials. (j) $CaF_2$ sample with two stacking faults (Ca-Ca bonds are drawn for clarity). (k) $CaF_2$ sample with two dislocations and stacking fault between them.

**2 Quantification of dislocation behaviour and fracture toughness induced by a weak magnetic field.**

The travelling distance of the dislocation slip trace with and without the influence of the magnetic field during micro-indentation was estimated from Fig. S2 and the results are displayed in Table S2. The change in dislocation density and fracture toughness under a weak magnetic field was subsequently calculated using Eqs. (21) and (23), and the results are also shown in Table S2.

**Table S2.** Travelling dislocation of utmost dislocation slip trace, dislocation density and fracture toughness induced by a magnetic field.

| Condition | NMF | MF |
|---|---|---|
| Travelling distance of utmost dislocation slip trace (µm) | 23.913±0.51 | 33.706±2.054 |
| Change in dislocation density ($\rho_{LB}/\rho_{L0}$) | 0.709 | |
| Change in fracture toughness ($K_{cB}/K_{c0}$) | 1.538 | |

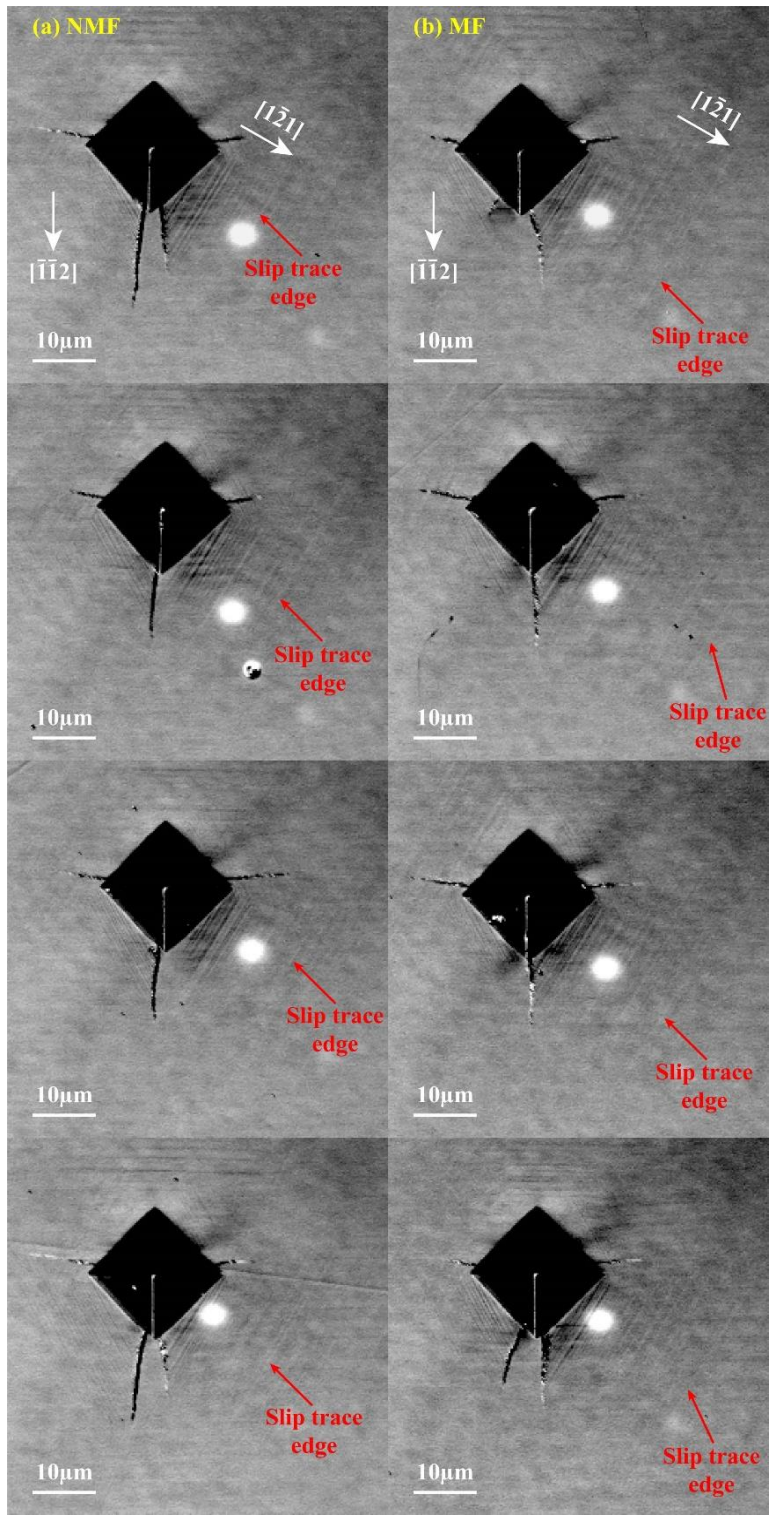

Fig. S2. Dislocation slip traces induced by the micro-indentation with a load of 0.5 N in the absence (a) and presence (b) of the magnetic field. (NMF: no magnetic field. MF: magnetic field).

**3. Magnetic field-induced ductile-brittle transition and its anisotropy in micro plunge-cutting**

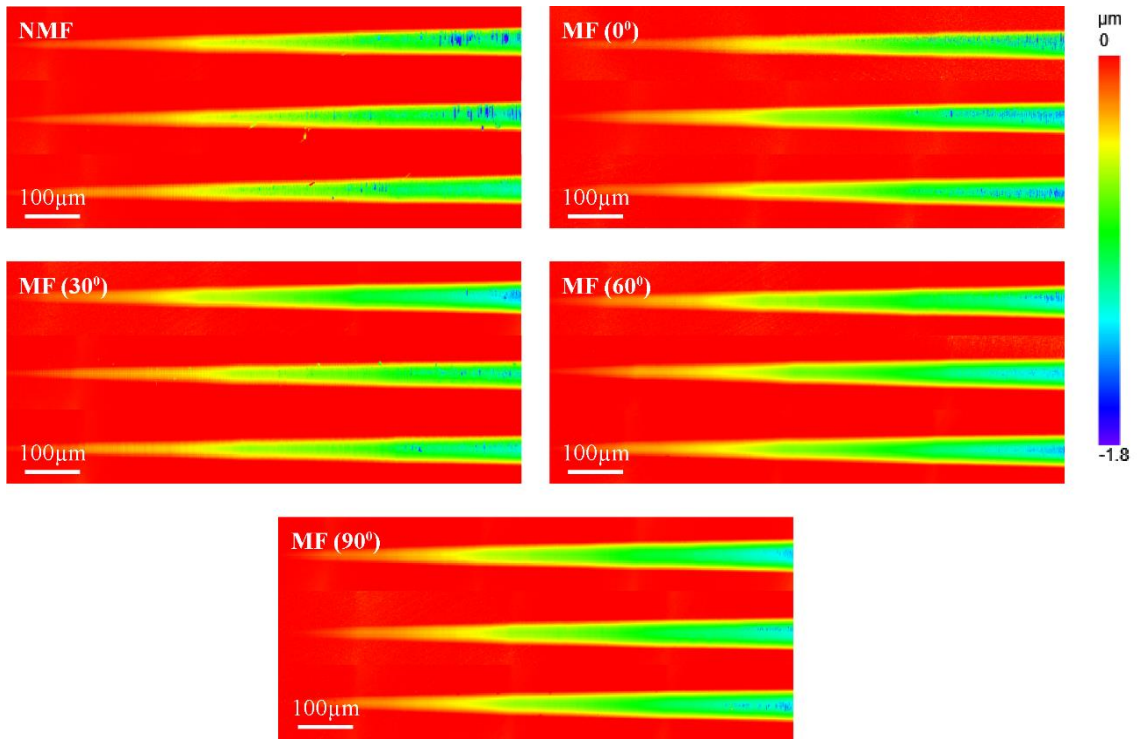

Fig. S3. Machined surface morphology of micro-grooves along [11$\bar{2}$] cutting direction with varying magnetic field directions.

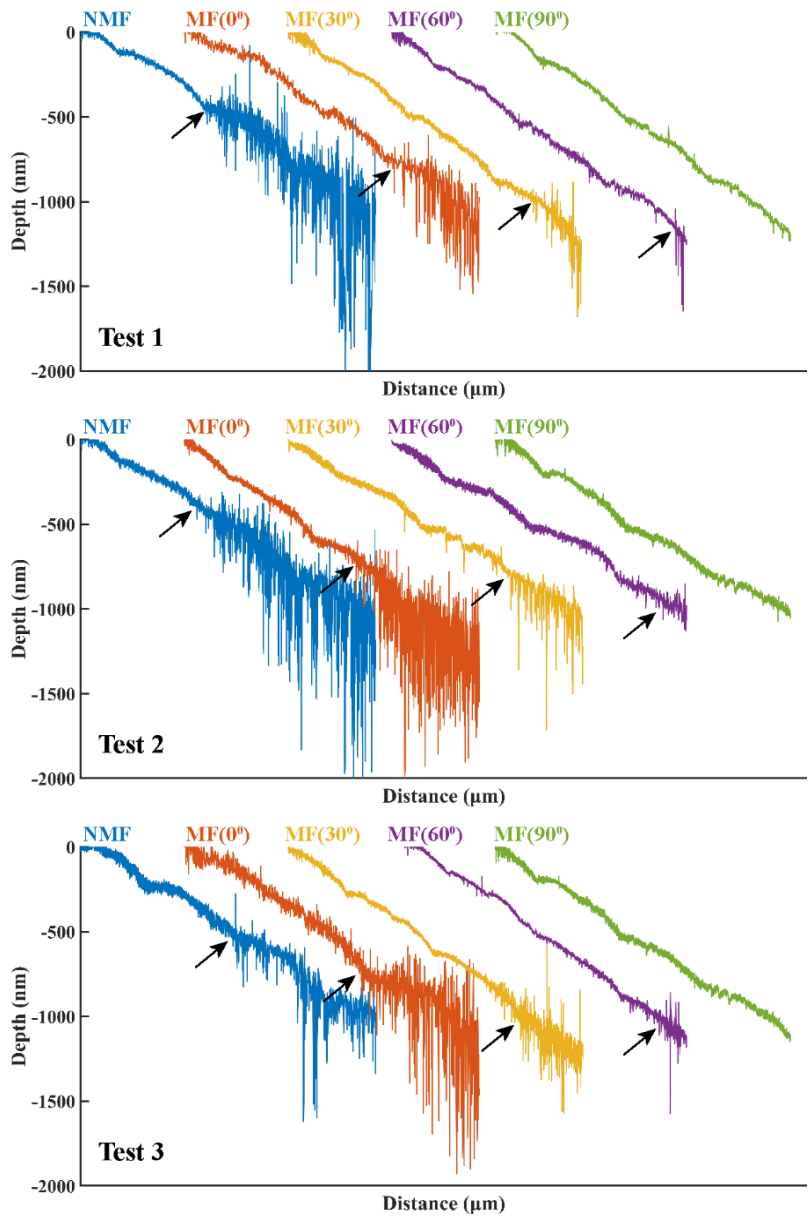

Fig. S4. Depth-distance curves of plunge cuttings along [11$\bar{2}$] cutting direction with varying magnetic field directions. (→: ductile–brittle transition point).

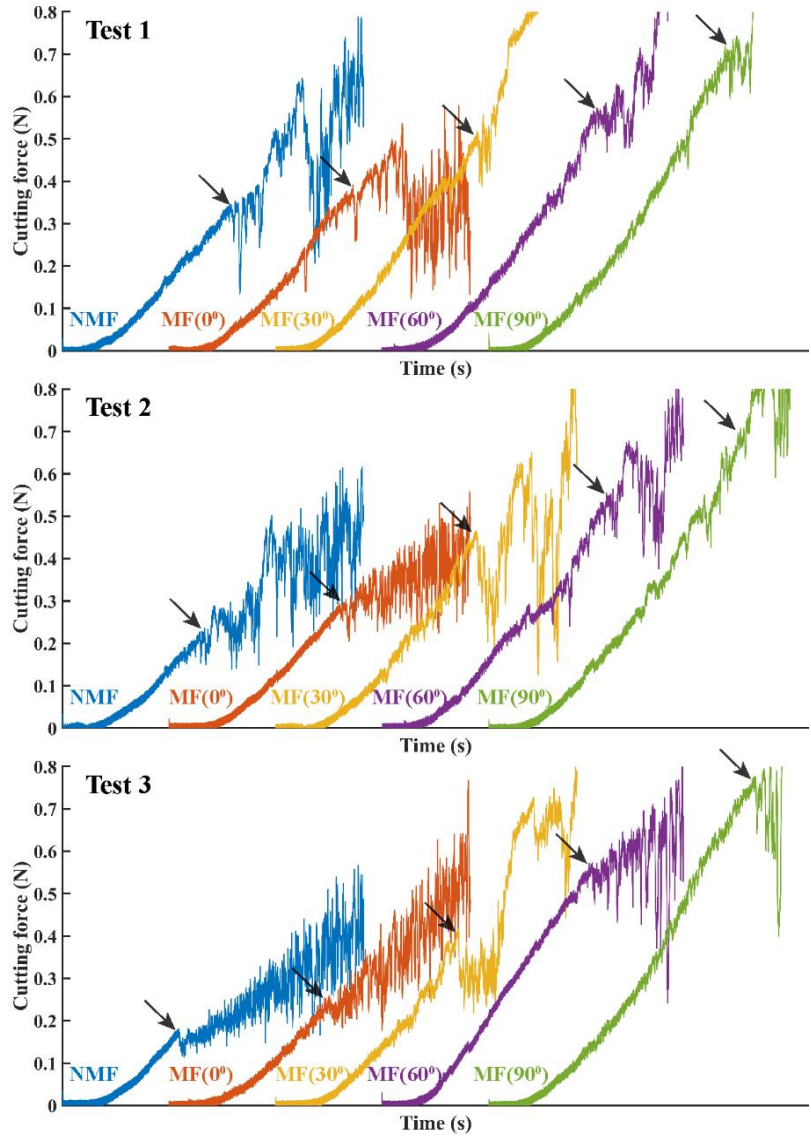

Fig. S5. Cutting force curves along [11$\bar{2}$] cutting direction at varying magnetic field directions. (→: ductile–brittle transition point).